\documentclass[aip,jcp,reprint,onecolumn]{revtex4}

\usepackage{amsmath}
\usepackage{mathtools}
\usepackage{amsbsy}
\usepackage{bm}
\usepackage{amssymb}
\usepackage{graphics}
\usepackage{graphicx}
\usepackage[colorlinks]{hyperref}
\usepackage{color}
\usepackage{subfig}
\usepackage{epstopdf}
\usepackage{esint}
\usepackage{multirow}

\allowdisplaybreaks

\def\bPsi{{\bm{\Psi}}}
\def\bg{{\mathbf{g}}}
\def\bk{{\mathbf{k}}}
\def\R{{\mathbb{R}}}

\def\bR{{\mathbf{R}}}
\def\bI{{\mathbf{I}}}
\def\bRJ{{\mathbf{R}_{J}}}
\def\bRJp{{\mathbf{R}_{J'}}}
\def\be{{\textbf{x}}}
\def\bS{{\mathbf{S}}}
\def\bg{{\mathbf{g}}}
\def\bF{{\mathbf{F}}}
\def\bQ{{\mathbf{Q}}}
\def\Fab{{F\mkern-3mu_{\alpha \beta}}}
\def\sab{{\sigma \mkern-2mu_{\alpha \beta}}}
\def\grada{{\nabla \mkern-6mu_{\alpha}}}
\def\gradb{{\nabla \mkern-6mu_{\beta}}}
\def\bk{{\mathbf{k}}}
\def\R{{\mathbb{R}}}
\def\occ{\textsl{g}}
\def\L{{\mathcal{L}}}
\def \gammaJl{\gamma_{_{Jl}}}
\def\tchins{{\tilde{\chi}_{_{Jlm}}}}

\def\pchins{{\chi_{_{J'lm}}}}
\def\tchis{{\tilde{\chi}^{*}_{_{Jlm}}}}

\def\pchis{{\chi^{*}_{_{J'lm}}}}

\def\bSmt{\mathbf{S}^{\scalebox{0.7}{\rm -T}}}
\def\bSt{\mathbf{S}^{\scalebox{0.7}{\rm T}}}
\def\bSi{\mathbf{S}^{\scalebox{0.7}{-1}}}
\def\bFmt{\mathbf{F}^{\scalebox{0.7}{\rm -T}}}

\def\bFi{\mathbf{F}^{\scalebox{0.7}{-1}}}

\begin{document}

\title{On the calculation of the stress tensor in real-space Kohn-Sham Density Functional Theory}
\author{Abhiraj Sharma and Phanish Suryanarayana}
\email[Email: ]{phanish.suryanarayana@ce.gatech.edu}
\affiliation{College of Engineering, Georgia Institute of Technology, GA 30332, USA}
\date{\today}
%%%%%%%%%%%%%%%%%%%%%%%%%%%%%%%%%%%%%%%%%%%%%%%%%%%%%%%%%%%%%%%%%%%%%%%%%%%%%%%%%%%%%%%%%%%%%%%%%%%%%%%%%%%%%%%%%%%%%%%%%%%%%%%%%%%%%%%%%%%%%%%%%%%%%%%%%%%%%%%%%%%%%%%%%%%%%%%%
\begin{abstract}
We present an accurate and efficient formulation of the stress tensor for real-space Kohn-Sham Density Functional Theory (DFT) calculations. Specifically, while employing a local formulation of the electrostatics, we derive a linear-scaling expression for the stress tensor that is applicable to simulations with unit cells of arbitrary symmetry, semilocal exchange-correlation functionals, and Brillouin zone integration. In particular, we rewrite the contributions arising from the self energy and the nonlocal pseudopotential energy to make them amenable to the  real-space finite-difference discretization, achieving up to three orders of magnitude improvement in the accuracy of the computed stresses. Using examples representative of static and dynamic calculations, we verify the accuracy and efficiency of the proposed formulation. In particular, we demonstrate high rates of convergence with spatial discretization, consistency between the computed energy and stress tensor, and very good agreement with reference planewave results. 
\end{abstract}
\maketitle
%%%%%%%%%%%%%%%%%%%%%%%%%%%%%%%%%%%%%%%%%%%%%%%%%%%%%%%%%%%%%%%%%%%%%%%%%%%%%%%%%%%%%%%%%%%%%%%%%%%%%%%%%%%%%%%%%%%%%%%%%%%%%%%%%%%%%%%%%%%%%%%%%%%%%%%%%%%%%%%%%%%%%%%%%%%%%%%
\section{Introduction}
Kohn-Sham Density Functional Theory (DFT) \cite{Hohenberg,Kohn1965} is an ab-initio method---technique relying on the first principles of quantum mechanics, without any empirical or experimental input---that is extensively used for understanding and predicting a wide variety of material properties. The tremendous popularity of DFT can be attributed to its highly favorable accuracy-to-cost ratio when compared to other such ab-initio approaches. In calculations of condensed matter systems using DFT, a basic quantity of interest in addition to the energy and atomic forces is the stress tensor, whose components represent derivatives of the energy density with respect to the different types of homogeneous strains that can be applied. The ability to compute the stress tensor components has a number of practical applications, ranging from the calculation of equilibrium lattice constants to the calculation of shear viscosity in ab initio molecular dynamics (AIMD). 

The derivation of the stress tensor in the context of DFT has its origins in the work of Slater \cite{Slater}, who obtained an expression for the pressure within the $X \alpha$ method. This was subsequently extended to any treatment of exchange and correlation by Janak \cite{Janak}. For pseudopotential DFT with local exchange-correlation functionals, the expression for the pressure was derived by Yin \cite{Yin}, who verified it using a planewave implementation. The complete stress tensor was  then derived by Nielsen and Martin \cite{Nielsen1,Nielsen2}, who also verified it using a planewave implementation. This formulation was later extended to the choice of semilocal exchange-correlation functionals by Corso and Resta \cite{StressGGA1994}, and to the linearized augmented plane wave (LAPW) and projector augmented-wave (PAW) methods by Thonhauser et al. \cite{Thonhauser2002} and Torrent et. al. \cite{PAWABINIT2008}, respectively \footnote{Though the implementation of the stress tensor in the context of PAW first appeared in the VASP code \cite{PAWKresse}, its derivation was not available until the work of Torrent et. al. \cite{PAWABINIT2008}.}. For atom-centered orbital bases, the stress tensors in pseudopotential and all-electron calculations were derived by Soler et. al. \cite{SIESTA} and Knuth et. al. \cite{Knuth2015}, respectively. The corresponding expressions within the finite-element discretization were recently derived by Motamarri and Gavini \cite{Motamarriconfigurationalforce}. 

Among the various discretizations that are systematically improvable and have a localized representation, the finite-difference method---all quantities of interest are expressed on a real-space grid---is perhaps the most mature and widely used to date in DFT \cite{beck2000rsmeth,saad2010esmeth}. In this approach, convergence is controlled by a single parameter and large-scale scalable parallel  implementations can be developed by virtue of the method's locality and freedom from communication-intensive transforms such as FFTs. Furthermore, a variety of boundary conditions can be accommodated, therefore enabling efficient and accurate treatment of finite, semi-infinite, as well as bulk 3D systems. Finally, linear-scaling methods can be developed \cite{pratapa2016spectral,MGMolref, suryanarayana2017sqdft}, by virtue of the discretization's locality and the nearsightedness of matter \cite{prodan2005nearsightedness,suryanarayana2017nearsightedness}. In view of these attractive features, significant advances have been made, whereby real-space finite-difference methods have been applied to systems containing thousands of atoms \cite{alamany2008fdlarge}, and have now outperformed established planewave codes \cite{ghosh2016sparc1,ghosh2016sparc2}, traditionally the method of choice in DFT. However, to the best of our knowledge, an expression for the stress tensor that is suitable for real-space calculations has not been derived heretofore, which provides the motivation for this work. 

In this work, we present an accurate and efficient formulation of the stress tensor in real-space DFT calculations. Specifically, while employing a local formulation of the electrostatics, we derive a linear-scaling expression for the stress tensor that is applicable to simulations consisting of unit cells with arbitrary symmetry, semilocal exchange-correlation functionals, and Brillouin zone integration. In particular, we rewrite the contributions to the stress tensor arising from the nonlocal pseudopotential energy and the self energy, making them amenable to the real-space method and thus achieving up to three orders of magnitude improvement in the accuracy of the computed stresses. We verify the accuracy and efficiency of the proposed formulation through examples representative of static and dynamic DFT calculations. In particular, we obtain high rates of convergence with spatial discretization, consistency between the computed energy and stress tensor, and very good agreement with reference planewave results.

The rest of this paper is organized as follows. First, we review the real-space formulation of Kohn-Sham DFT in Section~\ref{Section:formulation}. Using this framework, we derive an expression for the stress tensor in Section~\ref{Section:stress}. Next, we verify the accuracy and efficiency of the proposed formulation in Section~\ref{Section:implementation}. Finally, we provide concluding remarks in Section~\ref{Section:conclusion}.

%%%%%%%%%%%%%%%%%%%%%%%%%%%%%%%%%%%%%%%%%%%%%%%%%%%%%%%%%%%%%%%%%%%%%%%%%%%%%%%%%%%%%%%%%%%%%%%%%%%%%%%%%%%%%%%%%%%%%%%%%%%%%%%%%%%%%%%%%%%%%%%%%%%%%%%%%%%%%%%%%%%%%%%%%%%%%%%%%%           
\section{Real-space formulation of DFT}\label{Section:formulation}
\begin{figure}[h!]
\includegraphics[keepaspectratio=true,scale=0.8]{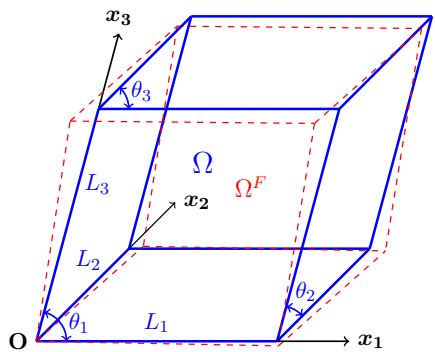}
\caption{\label{Fig:celldeformation} Unit cell $\Omega$ (solid blue lines) and its infinitesimally deformed version $\Omega^{F}$ (dashed red lines). The lattice vectors corresponding to $\Omega$ are $L_1 \bm{\hat{x}_1}$, $L_2 \bm{\hat{x}_2}$, and $L_3 \bm{\hat{x}_3}$, where $\bm{\hat{x}_1}$, $\bm{\hat{x}_2}$, and $\bm{\hat{x}_3}$ are the lattice unit vectors, with $\theta_1 =$ arccos$(\bm{\hat{x}_3}.\bm{\hat{x}_1})$, $\theta_2 =$ arccos$(\bm{\hat{x}_2}.\bm{\hat{x}_3})$, and $\theta_3 =$ arccos$(\bm{\hat{x}_1}.\bm{\hat{x}_2})$ representing the angles between them.}
\end{figure} 
Consider a unit cell $\Omega$ shown in Fig.~\ref{Fig:celldeformation} with lattice vectors $L_1 \bm{\hat{x}_1}$, $L_2 \bm{\hat{x}_2}$, and $L_3 \bm{\hat{x}_3}$, where $\bm{\hat{x}_1}$, $\bm{\hat{x}_2}$, and $\bm{\hat{x}_3}$ are the lattice unit vectors that are related to the Cartesian unit vectors $\bm{\hat{e}_1}$, $\bm{\hat{e}_2}$, and $\bm{\hat{e}_3}$ via the matrix $\mathbf{S}$, i.e., $\begin{bmatrix} \bm{\hat{x}_1} & \bm{\hat{x}_2} & \bm{\hat{x}_3} \end{bmatrix}^{\rm T} = \mathbf{S} \begin{bmatrix} \bm{\hat{e}_1} & \bm{\hat{e}_2} & \bm{\hat{e}_3} \end{bmatrix}^{\rm T}$. In this unit cell, let the nuclei be positioned at $\bR = \{\bR_1,\bR_2, \ldots ,\bR_N\}$ and contain a total of $N_e$ valence electrons. Neglecting spin and using the pseudopotential approximation, the free energy of the system in Kohn-Sham DFT \cite{Hohenberg,Kohn1965} at finite electronic temperature \cite{mermin1965thermal} can be written as 
\begin{equation} \label{Eqn:Energy}
\mathcal{F} (\bPsi,\bg,\phi,\bR) =  T_s(\bPsi,\bg) + E_{xc}(\rho,\bm{\nabla} \rho) +  E_{nl}(\bPsi,\bg,\bR) + E_{el}(\rho,\phi,\bR) - S(\bg) \,,
\end{equation}
where $T_s$ is the electronic kinetic energy, $E_{xc}$ is the exchange-correlation energy, $E_{nl}$ is the nonlocal pseudopotential energy, $E_{el}$ is the total electrostatic energy, $S$ is the electronic entropy energy, $\bPsi = \{\psi_1,\psi_2, \ldots ,\psi_{N_s}\}$ is the collection of orbitals with occupations $\bg = \{\occ_1,\occ_2, \ldots \occ_{N_s}\}$, $\phi$ is the electrostatic potential \cite{Pask2005,Phanish2010}, and $\rho$ is the electron density:
\begin{equation} \label{Eqn:electrondensity}
\rho(\be) = 2 \sum_{n=1}^{N_s} \fint_{BZ} \occ_n(\bk) {|\psi_n(\be,\bk)|}^2 \, \mathrm{d \bk} \,. \\
\end{equation} 
Above, $\bk$ denotes the wavevector and $\fint_{BZ}$ represents the volume average over the Brillouin zone. 

The electronic kinetic energy is of the form 
\begin{equation} \label{Eqn:electronickineticenergy}
T_s(\bPsi,\bg) = - \sum_{n=1}^{N_s} \fint_{BZ} \int_{\Omega} \occ_n(\bk) \psi_n^*(\be,\bk) \nabla^2 \psi_n(\be,\bk) \, \mathrm{d \be} \, \mathrm{d \bk} \,,
\end{equation}
where $\psi_n^*$ denotes the complex conjugate of $\psi_n$ and $\nabla^2 = \bm{\nabla}^{\scalebox{0.7}{\rm T}} \bm{\nabla}$, with $\bm{\nabla} = \bSi \big[\frac{\partial}{\partial x_1} \quad \frac{\partial}{\partial x_2} \quad \frac{\partial}{\partial x_3} \big]^{\scalebox{0.7}{\rm T}}$ being the gradient defined in the Cartesian coordinate system. The exchange-correlation energy within the semilocal generalized gradient approximation (GGA) \cite{langreth1980theory,langreth1981easily} can be expressed as 
\begin{equation} \label{Eqn:exchangecorrelationenergy}
E_{xc} (\rho,\bm{\nabla} \rho) = \int_{\Omega} \varepsilon_{xc} \big(\rho(\be),\bm{\nabla} \rho(\be)\big) \rho(\be) \, \mathrm{d \be} \,,\\
\end{equation} 
where $\varepsilon_{xc} = \varepsilon_x + \varepsilon_c$ is the sum of the exchange and correlation per particle of a uniform electron gas. The nonlocal pseudopotential energy within the Kleinman-Bylander \cite{kleinman1982efficacious} representation takes the form
\begin{equation}
E_{nl}(\bPsi,\bg,\bR) = 2 \sum_{n=1}^{N_s} \fint_{BZ} \occ_n(\bk) \sum_J \sum_{lm} \gammaJl \bigg | \int_{\Omega} \tchis(\be,\bRJ,\bk) \psi_n(\be,\bk) \, \mathrm{d\be} \bigg|^2 \, \mathrm{d \bk} \,,\\
\end{equation}
where the summation index $J$ runs over all atoms in $\Omega$, $lm$ runs over all azimuthal and magnetic quantum numbers, $\gammaJl$ is a normalization constant, and $\tchins$ are the Bloch-periodically mapped projectors, i.e., $\tchins= \sum_{J'} \pchins \, e^{-i \bk \cdot (\bRJ - \bRJp)}$. Here, the summation index $J'$ runs over $J^{th}$ atom and its periodic images, $\pchins$ is the corresponding projector, and $i = \sqrt{-1}$. The total electrostatic energy---locally reformulated \cite{Suryanarayana2014524,ghosh2016higher}, thus making it suitable for real-space calculations---can be written as 
\begin{equation}\label{Eqn:electrostaticenergy}
E_{el}(\rho,\phi,\bR) = - \frac{1}{8 \pi} \int_{\Omega} |\bm{\nabla} \phi(\be)|^2 \, \mathrm{d\be} + \int_{\Omega} \big(\rho(\be)+ b(\be,\bR) \big) \phi(\be) \, \mathrm{d\be} - E_{self}(\bR) + E_c(\bR) \,,   \\
\end{equation}
where $b = \sum_I b_I$ represents the total pseudocharge density of the nuclei, with $b_I$ being the pseudocharge density of the $I^{th}$ nucleus and the summation index $I$ running over all atoms in $\R^3$; $E_{self}(\bR) = \frac{1}{2}\sum_I \int_{\Omega} b_I(\be,\bR_I) V_I(\be,\bR_I) \, \mathrm{d\be}$ is the self energy associated with the pseudocharge densities, with $V_I$ being the pseudopotential of the $I^{th}$ nucleus; and $E_c$ corrects for the error in the repulsive energy when there is overlap of the pseudocharge densities (Appendix~\ref{subsection:electrostaticcorrection}). The electronic entropy energy arising due to the partial occupation of orbitals takes the form 
\begin{equation}\label{Eqn:electronicentropyenergy}
S(\bg) = -2 \sigma \sum_{n=1}^{N_{s}} \fint_{BZ} \Big(\occ_n(\bk) \log \occ_n(\bk) + \big(1-\occ_n(\bk)\big) \log \big(1-\occ_n(\bk)\big) \Big) \, \mathrm{d\bk} \,,
\end{equation}
where $\sigma$ is the smearing. 

The electronic ground state for a fixed position of nuclei is given by the solution of the constrained minimization problem \cite{Phanish2010,ghosh2016sparc2}
\begin{eqnarray}
& & \adjustlimits \min_{\bPsi,\, \bg} \max_{\phi} \mathcal{F}(\bPsi,\bg,\phi,\bR) \label{Eqn:groundstate} \\ 
 \text{s.t.} & & \int_{\Omega} \psi_m^*(\be,\bk) \psi_n(\be,\bk) \, \mathrm{d\be} = \delta _{m n} \,\, \forall \,\, \bk \quad \text{and} \quad 2 \sum_{n=1}^{N_s} \fint_{BZ} \occ_n(\bk) \, \mathrm{d \bk} = N_e \,,
\end{eqnarray}  
where $\delta_{mn}$ is the Kronecker delta function. In this variational problem, the orbitals are minimized over Bloch-periodic functions, i.e., $\psi_n(\be + \mathbf{L},\bk) = e^{i\bk.\mathbf{L}} \psi_n(\be,\bk)$ for every lattice vector $\mathbf{L}$ and Bloch wavevector $\bk$. The corresponding Euler-Lagrange equations take the form:
\begin{eqnarray}
\left( \mathcal{H} \equiv - \, \frac{1}{2} \nabla^2 + V_{xc} + \phi + V_{nl} \right)  \psi_n  &=&  \lambda_n \psi_n \,, \quad n=1,2, \ldots, N_s \,, \label{Eqn:eulerlagrange}\\
\occ_n(\bk) = \left(1+\exp\left(\frac{\lambda_n(\bk) - \lambda_f}{\sigma}\right)\right)^{-1} \,,& & \lambda_f \, \text{ is s.t. } \, 2 \sum_{n=1}^{N_s} \fint_{BZ} \occ_n(\bk) \, \mathrm{d \bk} = N_e \label{Eqn:orbitaloccupation} \,,\\
- \frac{1}{4 \pi} \nabla^2 \phi(\be,\bR) &=& \rho(\be) + b(\be,\bR)\,, \label{Eqn:poissonequation}
\end{eqnarray}
where $\mathcal{H}$ is the Hamiltonian with eigenfunctions $\psi_n$ and eigenvalues $\lambda_n$, $\lambda_f$ is the Fermi level, $V_{xc}$ is the exchange-correlation potential:
\begin{equation}
V_{xc} = \frac{\delta E_{xc}}{\delta \rho} = \varepsilon_{xc} + \rho \frac{\partial \varepsilon_{xc}}{\partial \rho} - \bm{\nabla}\cdot\left( \rho \frac{\partial \varepsilon_{xc}}{\partial (\bm{\nabla} \rho)} \right)\,, \label{Eqn:exchangecorrelationpotential} 
\end{equation}
and $V_{nl}$ is the nonlocal pseudopotential operator:
\begin{equation}
V_{nl} \psi_n = \sum_{J} \sum_{lm} \gammaJl \tchins \left(\int_{\Omega} \tchis(\be,\bRJ,\bk) \psi_n(\be,\bk)\, \mathrm{d\be} \right) \label{Eqn:nonlocalpseudopotential}\,.
\end{equation}
Once the electronic ground state has been determined, the Hellmann-Feynman stress tensor can be calculated, the expression for which we now derive in the framework described above.

%%%%%%%%%%%%%%%%%%%%%%%%%%%%%%%%%%%%%%%%%%%%%%%%%%%%%%%%%%%%%%%%%%%%%%%%%%%%%%%%%%%%%%%%%%%%%%%%%%%%%%%%%%%%%%%%%%%%%%%%%%%%%%%%%%%%%%%%%%%%%%%%%%%%%%%%%%%%%%%%%%%%%%%%%%%%%%%%
%%%%%%%%%%%%%%%%%%%%%%%%%%%%%%%%%%%%%%%%%%%%%%%%%%%%%%%%%%%%%%%%%%%%%%%%%%%%%%%%%%%%%%%%%%%%%%%%%%%%%%%%%%%%%%%%%%%%%%%%%%%%%%%%%%%%%%%%%%%%%%%%%%%%%%%%%%%%%%%%%%%%%%%%%%%%%%%%
%%%%%%%%%%%%%%%%%%%%%%%%%%%%%%%%%%%%%%%%%%%%%%%%%%%%%%%%%%%%%%%%%%%%%%%%%%%%%%%%%%%%%%%%%%%%%%%%%%%%%%%%%%%%%%%%%%%%%%%%%%%%%%%%%%%%%%%%%%%%%%%%%%%%%%%%%%%%%%%%%%%%%%%%%%%%%%%%

\section{Stress tensor in real-space DFT}\label{Section:stress}  
Consider an infinitesimal homogeneous deformation that maps the unit cell $\Omega$ to $\Omega^{F}$, as shown in Fig.~\ref{Fig:celldeformation}. Using $\bF$ to denote the deformation gradient in Cartesian coordinates, the stress tensor can be defined as 
\begin{equation} \label{Eqn:stressdefinition}
\sab = \frac{1}{|\Omega|} \frac{\partial \mathcal{L}^{F}(\bPsi,\bg,\phi,\bR \mkern-2mu^{F})}{\partial \Fab}\Bigg|_{\mathcal{G}} \,, \quad \alpha, \beta \in \{1, 2, 3\} \,,  
\end{equation}
where $|\Omega|$ is the measure of the unit cell \footnote{For systems are that are extended in three, two, and one dimensions, the measure of a unit cell is defined to be its volume, area, and length, respectively.},  the superscript $(.)^F$ is used to denote quantities after deformation---a notation adopted henceforth, $\mathcal{G}$ signifies the electronic ground state corresponding to the undeformed unit cell $\Omega$, i.e., at $\bF = \bI$, and the Lagrangian 
\begin{eqnarray}\label{Eqn:lagrangian} 
\L^F(\bPsi,\bg,\phi,\bR^F) &=& \mathcal{F}^F(\bPsi,\bg,\phi,\bR^F) -2 \sum_{mn} \fint_{BZ^F} \lambda_{mn}(\bk^F) \left(\int_{\Omega^F} \psi_m^*(\be^F,\bk^F) \psi_n(\be^F,\bk^F) \, \mathrm{d\be^F} - \delta _{m n} \right) \,\mathrm{d \bk^F} \nonumber \\
&-& \lambda_f \, \left(2 \sum_{n=1}^{N_s} \fint_{BZ^F} \occ_n(\bk^F) \, \mathrm{d \bk^F} - N_e\right)  \,.
\end{eqnarray}
Above, $\lambda_f$ and $\lambda_{mn} (m,n = 1,2, \cdots N_s)$ are the Lagrange multipliers used to enforce the constraint on the total number of electrons and the orthonormality of the orbitals, respectively.

In Sections~\ref{Subsection:kineticstress}-~\ref{Subsection:chargeconservationconstraintstress} below, we derive the contributions to the stress tensor arising from the various terms in $\L^F$, before presenting the expression for the total stress in Section~\ref{Subsection:completestresstensor}. In so doing, we will use a hat $\hat{(.)}$ to denote all ground state quantities, $\det(\bF)$ to denote the determinant of the matrix $\bF$, and $\grada$ ($\alpha \in 1,2,3$) to denote the $\alpha^{th}$ component of the gradient vector. In addition, we will use the relations: 
\begin{eqnarray}
\be^{F} = \bQ \be \,, \quad \bR \mkern-2mu^{F} = \bQ \bR \,, \quad \bk^{F} = \bQ^{\rm -T} \bk \,, \quad \bF^{-1} \approx 2 - \bF \,,
\end{eqnarray} 
where $\bQ = \bS^{\rm -T} \bF \bS^{\rm T}$. The final relation is a consequence of the deformation being infinitesimal in nature. 

%%%%%%%%%%%%%%%%%%%%%%%%%%%%%%%%%%%%%%%%%%%%%%%%%%%%%%%%%%%%%%%%%%%%%%%%%%%%%%%%%%%%%%%%%%%%%%%%%%%%%%%%%%%%%%%%%%%%%%%%%%%%%%%%%%%%%%%%%%%%%%%%%%%%%%%%%%%%%%%%%%%%%%%%%%%%%%%%
\subsection{Stress tensor contribution $\sigma \mkern-2mu^{T_s}$}\label{Subsection:kineticstress}
The contribution to the stress tensor arising from the electronic kinetic energy: 
\begin{eqnarray}
\sigma \mkern-2mu_{\alpha \beta}^{T_s} &=& \frac{\partial T_s^F(\bPsi,\bg)}{\partial \Fab} \Bigg|_{\mathcal{G}} \nonumber \\
&=& \frac{\partial}{\partial \Fab}\Bigg(-\sum_{n=1}^{N_s} \fint_{BZ^F} \int_{\Omega^F} \occ_n(\bk^{F}) \psi_n^*(\be^{F},\bk^{F}) \nabla^2 \psi_n(\be^{F},\bk^{F}) \, \mathrm{d \be^{F}} \, \mathrm{d \bk^{F}} \Bigg)\Bigg|_{\mathcal{G}} \nonumber \\
&=& \frac{\partial}{\partial \Fab}\Bigg(-\sum_{n=1}^{N_s} \fint_{BZ} \int_{\Omega} \occ_n(\mathbf{Q}^{\rm -T} \bk) \psi_n^*(\mathbf{Q} \be,\mathbf{Q} ^{\rm -T} \bk) \big[\bm{\nabla}^{\scalebox{0.7}{\rm T}}  \bFi \bFmt \bm{\nabla}\big] \psi_n(\mathbf{Q} \be,\mathbf{Q}^{\rm -T} \bk) \det (\bF) \, \mathrm{d \be} \, \mathrm{d \bk}\Bigg)\Bigg|_{\mathcal{G}} \nonumber \\
&=& A_1 + A_2 + A_3 + A_4 + A_5 \,, \label{Eqn:sigmaTs}
\end{eqnarray}
where
\begin{eqnarray}
A_1 &=& -\sum_{n=1}^{N_s} \fint_{BZ} \int_{\Omega} \frac{\partial \occ_n(\mathbf{Q}^{\rm -T} \bk)}{\partial \Fab} \bigg|_{\mathcal{G}} \hat{\psi}_n^*(\be,\bk) \nabla^2 \hat{\psi}_n(\be,\bk) \, \mathrm{d \be} \, \mathrm{d \bk}\nonumber \,,\\
A_2 &=& -\sum_{n=1}^{N_s} \fint_{BZ} \int_{\Omega} \hat{\occ}_n(\bk) \frac{\partial \psi_n^*(\mathbf{Q} \be,\mathbf{Q} ^{\rm -T} \bk)}{\partial\Fab} \bigg|_{\mathcal{G}} \nabla^2 \hat{\psi}_n(\be,\bk) \, \mathrm{d \be} \, \mathrm{d \bk}\nonumber \,, \\
A_3 &=& -\sum_{n=1}^{N_s} \fint_{BZ} \int_{\Omega} \hat{\occ}_n(\bk) \hat{\psi}_n^*(\be,\bk) \nabla^2 \frac{\partial \psi_n(\mathbf{Q} \be,\mathbf{Q} ^{\rm -T} \bk)}{\partial\Fab}\bigg|_{\mathcal{G}} \, \mathrm{d \be} \, \mathrm{d \bk}\nonumber \,,\\
A_4 &=& -\sum_{n=1}^{N_s} \fint_{BZ} \int_{\Omega} \hat{\occ}_n(\bk) \hat{\psi}_n^*(\be,\bk) \nabla^2 \hat{\psi}_n(\be,\bk) \frac{\partial \big(\det(\bF)\big)}{\partial \Fab} \bigg|_{\mathcal{G}}  \, \mathrm{d \be} \, \mathrm{d \bk}\nonumber \\  
    &=& -\sum_{n=1}^{N_s} \fint_{BZ} \int_{\Omega} \hat{\occ}_n(\bk) \hat{\psi}_n^*(\be,\bk) \nabla^2 \hat{\psi}_n(\be,\bk) \delta_{\alpha \beta} \, \mathrm{d \be} \, \mathrm{d \bk}\nonumber \,,\\
A_5 &=& -\sum_{n=1}^{N_s} \fint_{BZ} \int_{\Omega} \hat{\occ}_n(\bk) \hat{\psi}_n^*(\be,\bk) \frac{\partial \big(\bm{\nabla}^{\scalebox{0.7}{\rm T}} \bFi \bFmt \bm{\nabla} \big)}{\partial \Fab} \bigg|_{\mathcal{G}} \hat{\psi}_n(\be,\bk) \, \mathrm{d \be} \, \mathrm{d \bk} \nonumber \\
    &=& -\sum_{n=1}^{N_s} \fint_{BZ} \int_{\Omega} \hat{\occ}_n(\bk) \hat{\psi}_n^*(\be,\bk) \big[\mkern-5mu -\mkern-6mu \grada \gradb - \mkern-6mu \gradb \grada \big] \hat{\psi}_n(\be,\bk) \, \mathrm{d \be} \, \mathrm{d \bk} \nonumber \\
    &=& -2 \,\sum_{n=1}^{N_s} \fint_{BZ} \int_{\Omega} \hat{\occ}_n(\bk) \grada \hat{\psi}_n^*(\be,\bk)  \gradb \hat{\psi}_n(\be,\bk) \, \mathrm{d \be} \, \mathrm{d \bk} \nonumber \,.
\end{eqnarray}
The last equality in $A_5$ is obtained via integration by parts, performed to reduce the number of derivative evaluations and circumvent the need for mixed derivatives, which are typically more costly to evaluate within the real-space method \cite{natan2008real,AbhirajKP}. 
%%%%%%%%%%%%%%%%%%%%%%%%%%%%%%%%%%%%%%%%%%%%%%%%%%%%%%%%%%%%%%%%%%%%%%%%%%%%%%%%%%%%%%%%%%%%%%%%%%%%%%%%%%%%%%%%%%%%%%%%%%%%%%%%%%%%%%%%%%%%%%%%%%%%%%%%%%%%%%%%%%%%%%%%%%%%%%%%
\subsection{Stress tensor contribution $\sigma \mkern-2mu ^{E_{xc}}$}\label{Subsection:exchangecorrelationstress}
The contribution to the stress tensor arising from the exchange-correlation energy: 
\begin{eqnarray}
\sigma \mkern-2mu ^{E_{xc}}_{\alpha \beta} &=& \frac{\partial E_{xc}^F(\rho,\bm{\nabla} \rho)}{\partial\Fab} \Bigg|_{\mathcal{G}} \nonumber \\
&=& \frac{\partial}{\partial\Fab}\Bigg( \int_{\Omega^F} \varepsilon_{xc} \big(\rho(\be^{F}),\bm{\nabla} \rho(\be^{F}) \big) \rho(\be^{F}) \, \mathrm{d \be^{F}} \Bigg) \Bigg|_{\mathcal{G}} \nonumber \\
&=& \frac{\partial}{\partial\Fab}\Bigg( \int_{\Omega} \varepsilon_{xc}\big(\rho(\mathbf{Q} \be),\bFmt \bm{\nabla} \rho(\mathbf{Q} \be) \big) \rho(\mathbf{Q} \be) \det(\bF) \, \mathrm{d \be} \Bigg) \Bigg|_{\mathcal{G}} \nonumber \\
&=& B_1 + B_2 + B_3 + B_4 + B_5 \,, \label{Eqn:sigmaExc}
\end{eqnarray}
where
\begin{eqnarray}
B_1 &=& 2 \, \sum_{n=1}^{N_s} \fint_{BZ} \int_\Omega V_{xc}\big(\hat{\rho}(\be),\bm{\nabla} \hat{\rho}(\be)\big) \frac{\partial \occ_n(\mathbf{Q}^{\rm -T} \bk)}{\partial\Fab} \bigg|_{\mathcal{G}} {\big|\hat{\psi}_n(\be,\bk)\big|}^2 \, \mathrm{d \be} \nonumber \,, \\
B_2 &=& 2 \, \sum_{n=1}^{N_s} \fint_{BZ} \int_\Omega V_{xc}\big(\hat{\rho}(\be),\bm{\nabla} \hat{\rho}(\be)\big) \hat{\occ}_n(\bk) \frac{\partial \psi_n^*(\mathbf{Q} \be,\mathbf{Q} ^{\rm -T} \bk)}{\partial\Fab} \bigg|_{\mathcal{G}} \hat{\psi}_n(\be,\bk) \, \mathrm{d \be} \nonumber \,, \\
B_3 &=& 2 \, \sum_{n=1}^{N_s} \fint_{BZ} \int_\Omega V_{xc}\big(\hat{\rho}(\be),\bm{\nabla} \hat{\rho}(\be)\big) \hat{\occ}_n(\bk) \hat{\psi}_n^*(\be,\bk) \frac{\partial \psi_n(\mathbf{Q} \be,\mathbf{Q} ^{\rm -T} \bk)}{\partial\Fab} \bigg|_{\mathcal{G}} \, \mathrm{d \be} \nonumber \,, \\
B_4 &=& \int_\Omega \varepsilon_{xc} \big(\hat{\rho}(\be),\bm{\nabla} \hat{\rho}(\be)\big) \hat{\rho}(\be) \frac{\partial \big(\det(\bF)\big)}{\partial \Fab} \bigg|_{\mathcal{G}} \, \mathrm{d \be} \nonumber \\ 
    &=& \int_\Omega \varepsilon_{xc} \big(\hat{\rho}(\be),\bm{\nabla} \hat{\rho}(\be)\big) \hat{\rho}(\be) \delta_{\alpha \beta} \, \mathrm{d \be} \nonumber\\
    &=& \delta_{\alpha \beta} E_{xc}(\hat{\rho}, \bm{\nabla} \hat{\rho}) \nonumber \,, \\  
B_5 &=& \int_\Omega \hat{\rho}(\be) \frac{\partial \varepsilon_{xc} \big(\hat{\rho}(\be),\bm{\nabla} \hat{\rho}(\be) \big)}{\partial \big(\bm{\nabla} \hat{\rho}(\be) \big)} \frac{\partial \bFmt} {\partial \Fab} \bigg|_{\mathcal{G}} \bm{\nabla} \hat{\rho}(\be) \, \mathrm{d \be} \nonumber \\
    &=& - \int_\Omega \hat{\rho}(\be) \frac{\partial \varepsilon_{xc}\big(\hat{\rho}(\be),\bm{\nabla} \hat{\rho}(\be) \big)}{\partial \big(\gradb \hat{\rho}(\be)\big)} \grada \hat{\rho}(\be) \, \mathrm{d \be} \nonumber \,.     
\end{eqnarray}
In obtaining the terms $B_1$, $B_2$, and $B_3$, we have used the relation:
\begin{eqnarray}
\frac{\partial \rho(\mathbf{Q} \be)}{\partial\Fab} &=& 2\, \sum_{n=1}^{N_s} \fint_{BZ} \Bigg( \frac{\partial \occ_n(\mathbf{Q}^{\rm -T} \be)}{\partial\Fab} \, {\big|\psi_n(\mathbf{Q} \be,\mathbf{Q} ^{\rm -T} \bk)\big|}^2 + \occ_n(\mathbf{Q} ^{\rm -T} \bk) \frac{\partial \psi^*_n(\mathbf{Q} \be,\mathbf{Q} ^{\rm -T} \bk)}{\partial\Fab} \psi_n(\mathbf{Q} \be,\mathbf{Q} ^{\rm -T} \bk) \nonumber \\
&+& \occ_n(\mathbf{Q} ^{\rm -T} \bk) \psi^*_n(\mathbf{Q} \be,\mathbf{Q} ^{\rm -T} \bk) \frac{\partial \psi_n(\mathbf{Q} \be,\mathbf{Q} ^{\rm -T} \bk)}{\partial\Fab} \Bigg)\, \mathrm{d \bk} \,. \label{Eqn:DerRhoFab}
\end{eqnarray}
%%%%%%%%%%%%%%%%%%%%%%%%%%%%%%%%%%%%%%%%%%%%%%%%%%%%%%%%%%%%%%%%%%%%%%%%%%%%%%%%%%%%%%%%%%%%%%%%%%%%%%%%%%%%%%%%%%%%%%%%%%%%%%%%%%%%%%%%%%%%%%%%%%%%%%%%%%%%%%%%%%%%%%%%%%%%%%%%
\subsection{Stress tensor contribution $\sigma \mkern-2mu ^{E_{nl}}$}\label{Subsection:nonlocalpseudopotentialstress}
The contribution to the stress tensor arising from the nonlocal pseudopotential energy:
\begin{eqnarray}
\sigma \mkern-2mu ^{E_{nl}}_{\alpha\beta} &=& \frac{\partial E_{nl}^F(\bPsi,\bg,\bR \mkern-2mu^{F})}{\partial\Fab} \Bigg|_{\mathcal{G}} \nonumber \\
&=& \frac{\partial}{\partial\Fab}\Bigg( 2 \sum_{n=1}^{N_s} \fint_{BZ^F} \occ_n(\bk^{F}) \sum_J \sum_{lm} \gammaJl \bigg | \int_{\Omega^F} \tchis(\be^{F},\bR_{J}^{\mkern-2mu{F}},\bk^{F}) \psi_n(\be^{F},\bk^{F}) \, \mathrm{d \be^{F}} \bigg|^2 \, \mathrm{d \bk^{F}} \Bigg)\Bigg|_{\mathcal{G}} \nonumber \\  
&=& \frac{\partial}{\partial\Fab}\Bigg( 2 \sum_{n=1}^{N_s} \fint_{BZ} \occ_n(\mathbf{Q} ^{\rm -T} \bk) \sum_J \sum_{lm} \gammaJl \bigg | \int_{\Omega} \tchis(\mathbf{Q} \be,\mathbf{Q} \bR _{J},\mathbf{Q} ^{\rm -T} \bk) \psi_n(\mathbf{Q} \be,\mathbf{Q} ^{\rm -T} \bk) \det(\bF)\, \mathrm{d\be} \bigg|^2 \, \mathrm{d \bk} \Bigg)\Bigg|_{\mathcal{G}} \nonumber \\  
&=& C_1 + C_2 + C_3 + C_4 + C_5 \,, \label{Eqn:sigmaEnl} 
\end{eqnarray}
where
\begin{eqnarray}
C_1 &=& 2 \sum_{n=1}^{N_s} \fint_{BZ} \frac{\partial \occ_n(\mathbf{Q} ^{\rm -T} \bk)}{\partial\Fab} \bigg|_{\mathcal{G}} \sum_J \sum_{lm} \gammaJl \bigg|\int_\Omega \tchis(\be,\bRJ,\bk) \hat{\psi}_n(\be,\bk) \, \mathrm{d \be} \bigg|^2 \, \mathrm{d \bk} \nonumber \,, \\ 
C_2 &=& 2 \sum_{n=1}^{N_s} \fint_{BZ} \hat{\occ}_n(\bk) \sum_J \sum_{lm} \gammaJl \bigg(\int_\Omega \tchis(\be,\bRJ,\bk) \hat{\psi}_n(\be,\bk)\, \mathrm{d \be}\bigg) \bigg(\int_\Omega \tchins(\be,\bRJ,\bk) \frac{\partial \psi^{*}_n(\mathbf{Q} \be,\mathbf{Q} ^{\rm -T} \bk)}{\partial\Fab} \bigg|_{\mathcal{G}} \, \mathrm{d \be} \bigg) \, \mathrm{d \bk} \nonumber \,, \\ 
C_3 &=& 2 \sum_{n=1}^{N_s} \fint_{BZ} \hat{\occ}_n(\bk) \sum_J \sum_{lm} \gammaJl \bigg(\int_\Omega \tchis(\be,\bRJ,\bk) \frac{\partial \psi_n(\mathbf{Q} \be,\mathbf{Q} ^{\rm -T} \bk)}{\partial\Fab} \bigg|_{\mathcal{G}}\, \mathrm{d \be}\bigg) \bigg(\int_\Omega \tchins(\be,\bRJ,\bk) \hat{\psi}^*_n(\be,\bk) \, \mathrm{d \be} \bigg) \, \mathrm{d \bk} \nonumber \,, \\
C_{4} &=& 4 \sum_{n=1}^{N_s} \fint_{BZ} \hat{\occ}_n(\bk) \sum_J \sum_{lm} \gammaJl \bigg(\int_{\Omega} \tchis(\be,\bRJ,\bk) \hat{\psi}_n(\be,\bk) \frac{\partial \big(\det(\bF)\big)}{\partial\Fab} \bigg|_{\mathcal{G}} \, \mathrm{d\be} \bigg) \bigg(\int_{\Omega} \tchins(\be,\bRJ,\bk) \hat{\psi}^{*}_n(\be,\bk) \, \mathrm{d\be} \bigg) \, \mathrm{d \bk} \nonumber \\
       &=& 4 \sum_{n=1}^{N_s} \fint_{BZ} \hat{\occ}_n(\bk) \sum_J \sum_{lm} \gammaJl \bigg(\int_{\Omega} \tchis(\be,\bRJ,\bk) \hat{\psi}_n(\be,\bk) \delta_{\alpha \beta} \, \mathrm{d\be} \bigg) \bigg(\int_{\Omega} \tchins(\be,\bRJ,\bk) \hat{\psi}^{*}_n(\be,\bk) \, \mathrm{d\be} \bigg) \, \mathrm{d \bk}  \nonumber \\ 
       &=& 2 \delta_{\alpha \beta} E_{nl}(\hat{\bPsi},\hat{\bg},\bR) \nonumber \,, \\       
C_5 &=& 4 \sum_{n=1}^{N_s} \fint_{BZ} \hat{\occ}_n(\bk) \sum_J \sum_{lm} \gammaJl \Re \Bigg[ \bigg(\int_\Omega  \frac{\partial \tchis(\mathbf{Q} \be,\mathbf{Q} \bR _{J},\mathbf{Q} ^{\rm -T} \bk)}{\partial \Fab} \Bigg|_{\mathcal{G}} \hat{\psi}_n(\be,\bk)  \, \mathrm{d \be} \bigg) \nonumber \bigg(\int_\Omega \tchins(\be,\bRJ,\bk) \hat{\psi}^*_n(\be,\bk) \, \mathrm{d \be}\bigg) \Bigg] \, \mathrm{d \bk} \nonumber \\
&=& 4 \sum_{n=1}^{N_s} \fint_{BZ} \hat{\occ}_n(\bk) \sum_J \sum_{lm} \gammaJl \Re \Bigg[ \bigg(\sum_{J'} \int_\Omega \grada \pchis(\be,\bRJp) e^{i \bk \cdot (\bRJ - \bRJp)} {\big(\bSt \be - \bSt \bRJp \big)\mkern-4mu}_\beta \hat{\psi}_n(\be,\bk)  \, \mathrm{d \be} \bigg) \nonumber \\
& \times & \bigg(\int_\Omega \tchins(\be,\bRJ,\bk) \hat{\psi}^*_n(\be,\bk) \, \mathrm{d \be}\bigg) \Bigg] \, \mathrm{d \bk} \nonumber \\
&=& -4 \sum_{n=1}^{N_s} \fint_{BZ} \hat{\occ}_n(\bk) \sum_J \sum_{lm} \gammaJl \Re \Bigg[ \bigg(\sum_{J'} \int_\Omega \pchis(\be,\bRJp) e^{i \bk \cdot (\bRJ - \bRJp)} {\big(\bSt \be - \bSt \bRJp \big)\mkern-4mu}_\beta \grada \hat{\psi}_n(\be,\bk)  \, \mathrm{d \be} \bigg) \nonumber \\
& \times & \bigg(\int_\Omega \tchins(\be,\bRJ,\bk) \hat{\psi}^*_n(\be,\bk) \, \mathrm{d \be}\bigg) \Bigg] \, \mathrm{d \bk} - 2 \delta_{\alpha \beta} E_{nl}(\hat{\bPsi},\hat{\bg},\bR) \nonumber \,.
\end{eqnarray}
In deriving the expression for $C_5$, the second equality is obtained as follows:
\begin{eqnarray}
\frac{\partial \tchis(\mathbf{Q} \be,\mathbf{Q} \bR _{J},\mathbf{Q} ^{\rm -T} \bk)}{\partial \Fab} \Bigg|_{\mathcal{G}} & = & \sum_{J'} \bigg(\frac{\partial \pchis(\mathbf{Q} \be,\mathbf{Q} \bR _{J'})}{\partial (\mathbf{Q} \be)} \frac{\partial (\mathbf{Q} \be)}{\partial \Fab} + \frac{\partial \pchis(\mathbf{Q} \be,\mathbf{Q} \bR _{J'})}{\partial (\mathbf{Q} \bR _{J'})} \frac{\partial (\mathbf{Q} \bR_{J'})}{\partial \Fab} \bigg) \bigg|_{\mathcal{G}}  e^{i \bk \cdot (\bRJ - \bRJp)}  \nonumber \\
& = & \sum_{J'} \bigg[\grada \pchis(\be,\bRJp) {\big(\bSt \be \big)\mkern-4mu}_\beta + {\bigg(\frac{\partial \pchis(\be,\bRJp)}{\partial \bRJp} \bSmt \bigg)\mkern-6mu}_\alpha {\big(\bSt \bRJp \big)\mkern-4mu}_\beta \bigg] \, e^{i \bk \cdot (\bRJ - \bRJp)} \, \nonumber \\
%& = &  \sum_{J'} {\bigg(\frac{\partial \pchis(\be,\bRJp)}{\partial \bRJp} \bSmt \bigg)\mkern-6mu}_\alpha  e^{i \bk \cdot (\bRJ - \bRJp)} {\big(\bSt \bRJp - \bSt \be \big)\mkern-4mu}_\beta \,, 
& = &  \sum_{J'} \grada \pchis(\be,\bRJp)  e^{i \bk \cdot (\bRJ - \bRJp)} {\big(\bSt \be - \bSt \bRJp \big)\mkern-4mu}_\beta \,, 
\end{eqnarray}
where the final equality is obtained by using the relation $\pchis(\be,\bRJp) = \pchis(\bSt \be - \bSt \bRJp)$. In deriving the expression for $C_5$, the third equality is obtained as follows:
\begin{eqnarray} 
\chi_{\alpha \beta}^{\sigma}  & = & \sum_{J'} \int_\Omega \grada \pchis(\be,\bRJp) e^{i \bk \cdot (\bRJ - \bRJp)} {\big(\bSt \be - \bSt \bRJp \big)\mkern-4mu}_\beta \hat{\psi}_n(\be,\bk) \, \mathrm{d \be} \nonumber \\
& = & \sum_{J'} \int_\Omega {\bigg(\frac{\partial \pchis(\be,\bRJp)}{\partial \bRJp} \bSmt \bigg)\mkern-6mu}_\alpha  e^{i \bk \cdot (\bRJ - \bRJp)} {\big(\bSt \bRJp - \bSt \be \big)\mkern-4mu}_\beta \hat{\psi}_n(\be,\bk) \, \mathrm{d \be} \nonumber \\
& = & \sum_{J'} \int_\Omega {\bigg(\frac{\partial}{\partial \bRJp} \bSmt \bigg)\mkern-6mu}_\alpha \bigg[\pchis(\be,\bRJp) {\big(\bSt \bRJp - \bSt \be \big)\mkern-4mu}_\beta e^{i \bk \cdot (\bRJ - \bRJp)} \hat{\psi}_n(\be,\bk) \bigg] \, \mathrm{d \be} \nonumber \\
& - & \delta_{\alpha \beta} \sum_{J'} \int_\Omega \pchis(\be,\bRJp)  e^{i \bk \cdot (\bRJ - \bRJp)} \hat{\psi}_n(\be,\bk) \, \mathrm{d \be} \nonumber \,, \\
& = & - \sum_{J'} \int_\Omega {\bigg(\frac{\partial}{\partial \bRJp} \bSmt \bigg)\mkern-6mu}_\alpha \bigg[\pchis(\bm{\eta}) {(\bSt \bm{\eta}) \mkern-3mu}_\beta e^{i \bk \cdot (\bRJ - \bRJp)} \hat{\psi}_n(\bm{\eta} + \bRJp,\bk) \bigg] \, \mathrm{d \bm{\eta}} - \delta_{\alpha \beta} \, \int_\Omega \tchis(\be,\bRJ,\bk) \hat{\psi}_n(\be,\bk) \, \mathrm{d \be} \nonumber \\
& = & \sum_{J'} \int_\Omega \pchis(\be,\bRJp) {\big(\bSt \bRJp - \bSt \be \big)\mkern-4mu}_\beta \, e^{i \bk \cdot (\bRJ - \bRJp)} \grada \hat{\psi}_n(\be,\bk) \, \mathrm{d \be} - \delta_{\alpha \beta} \int_\Omega \tchis(\be,\bRJ,\bk) \hat{\psi}_n(\be,\bk) \, \mathrm{d \be}  \,, \label{Eqn:NonlocalReformulation}
\end{eqnarray}
where the second equality is obtained by using the relation $\pchis(\be,\bRJp) = \pchis(\bSt \be - \bSt \bRJp)$, the third equality is obtained by using the chain-rule, the fourth equality is obtained by making the substitution $\be = \bm{\eta} + \bRJp$, and the final equality is obtained by taking the derivative with respect to $\bRJp$ and making the substitution $\bm{\eta} = \be - \bRJp$. 

The above reformulation is motivated by the fact that the original expression for $\chi_{\alpha \beta}^{\sigma}$ contains derivatives of the projectors, which are themselves highly localized and rapidly varying. Therefore, adopting the procedure previously used for the reformulation of the nonlocal component of the atomic forces \cite{hirose2005first,andrade2015real,ghosh2016sparc2}, we have transferred the derivative on the nonlocal projectors (with respect to atomic position) to the orbitals (with respect to space). Since the orbitals are typically more smooth than the projectors, the accuracy of the stress tensor is significantly improved due to this reformulation, as demonstrated in Appendix~\ref{subsection:stresswithoutnonlocalreformulation}. 

%%%%%%%%%%%%%%%%%%%%%%%%%%%%%%%%%%%%%%%%%%%%%%%%%%%%%%%%%%%%%%%%%%%%%%%%%%%%%%%%%%%%%%%%%%%%%%%%%%%%%%%%%%%%%%%%%%%%%%%%%%%%%%%%%%%%%%%%%%%%%%%%%%%%%%%%%%%%%%%%%%%%%%%%%%%%%%%%

\subsection{Stress tensor contribution $\sigma \mkern-2mu ^{E_{el}}$}\label{Subsection:electrostaticstress}
The contribution to the stress tensor arising from the total electrostatic energy: 
\begin{eqnarray}
\sigma \mkern-2mu ^{E_{el}}_{\alpha \beta} &=& \frac{\partial E_{el}^F(\rho,\bR \mkern-2mu^{F},\phi)}{\partial\Fab} \Bigg|_{\mathcal{G}} \nonumber \\
&=& \frac{\partial}{\partial\Fab} \Bigg(- \frac{1}{8 \pi} \int_{\Omega^F} \Big|\bm{\nabla} \phi(\be^{F})\Big|^2 \, \mathrm{d \be^{F}} + \int_{\Omega^F} \big(\rho(\be^{F}) + b(\be^{F},\bR \mkern-2mu^{F}) \big) \phi(\be^{F}) \, \mathrm{d \be^{F}} - E_{self}(\bR \mkern-2mu^{F}) + E_c(\bR \mkern-2mu^{F}) \Bigg)\Bigg|_{\mathcal{G}} \nonumber \\ 
&=& \frac{\partial}{\partial\Fab} \Bigg(- \frac{1}{8 \pi} \int_{\Omega} \Big|\bFmt \bm{\nabla} \phi(\mathbf{Q} \be)\Big|^2 \det(\bF) \, \mathrm{d \be} + \int_{\Omega} \big(\rho(\mathbf{Q} \be) + b(\mathbf{Q} \be, \mathbf{Q} \bR) \big) \phi(\mathbf{Q} \be) \det(\bF) \, \mathrm{d\be} - E_{self}(\mathbf{Q} \bR) + E_c(\mathbf{Q} \bR )\Bigg)\Bigg|_{\mathcal{G}} \nonumber \\ 
&=& D_1 + D_2 + D_3 + D_4 + D_5 - \sigma \mkern-2mu ^{E_{self}}_{\alpha\beta} + \sigma \mkern-2mu ^{E_{c}}_{\alpha\beta} \,, \label{Eqn:sigmaEel}
\end{eqnarray}
where
\begin{eqnarray} 
D_1 &=& 2 \sum_{n=1}^{N_s} \fint_{BZ} \int_\Omega \hat{\phi}(\be,\bR) \frac{\partial \occ_n(\mathbf{Q} ^{\rm -T} \bk)}{\partial\Fab} \bigg|_{\mathcal{G}} \, {\big|\hat{\psi}_n(\be,\bk)\big|}^2 \, \mathrm{d \be} \nonumber \,, \\
D_2 &=& 2 \sum_{n=1}^{N_s} \fint_{BZ} \int_\Omega \hat{\phi}(\be,\bR) \hat{\occ}_n(\bk) \frac{\partial \psi_n^*(\mathbf{Q} \be,\mathbf{Q} ^{\rm -T} \bk)}{\partial\Fab} \bigg|_{\mathcal{G}} \hat{\psi}_n(\be,\bk) \, \mathrm{d \be} \nonumber \,, \\
D_3 &=& 2 \sum_{n=1}^{N_s} \fint_{BZ} \int_\Omega \hat{\phi}(\be,\bR) \hat{\occ}_n(\bk) \hat{\psi}_n^*(\be,\bk) \frac{\partial \psi_n(\mathbf{Q} \be,\mathbf{Q} ^{\rm -T} \bk)}{\partial\Fab} \bigg|_{\mathcal{G}} \, \mathrm{d \be} \nonumber \,, \\
D_4 &=& \int_\Omega \bigg(- \frac{1}{8 \pi} \big|\bm{\nabla} \hat{\phi}(\be,\bR)\big|^2 + \big(\hat{\rho}(\be) + b(\be,\bR) \big) \hat{\phi}(\be,\bR) \bigg) \frac{\partial \big(\det(\bF)\big)}{\partial \Fab} \bigg|_{\mathcal{G}} \, \mathrm{d \be} \nonumber \\   
    &=& \int_\Omega \frac{1}{2} \big(\hat{\rho}(\be) + b(\be,\bR) \big) \hat{\phi}(\be,\bR) \delta_{\alpha \beta} \, \mathrm{d \be} \nonumber \,, \\   
D_5 &=& \int_\Omega \bigg(-\frac{1}{8 \pi} \frac{\partial }{\partial \Fab} \Big[\bm{\nabla}^{\rm T} \bFi \phi(\mathbf{Q} \be) \bFmt \bm{\nabla} \phi(\mathbf{Q} \be)\Big] \bigg|_{\mathcal{G}} + \big(\hat{\rho}(\be) + b(\be,\bR) \big) \frac{\partial \phi(\mathbf{Q} \be)}{\partial \Fab} \bigg|_{\mathcal{G}} + \frac{\partial b(\mathbf{Q} \be,\mathbf{Q} \bR)}{\partial \Fab}\bigg|_{\mathcal{G}} \hat{\phi}(\be,\bR) \bigg) \, \mathrm{d \be} \nonumber \\
   &=& \int_\Omega \bigg(\frac{1}{4 \pi} \grada \hat{\phi}(\be,\bR) \gradb \hat{\phi}(\be,\bR) + \bigg(\frac{1}{4 \pi} \nabla^2 \hat{\phi}(\be,\bR) \frac{\partial \phi(\mathbf{Q} \be)}{\partial\Fab} + \big(\hat{\rho}(\be) + b(\be,\bR)\big) \frac{\partial \phi(\mathbf{Q} \be)}{\partial\Fab} \bigg)\bigg|_{\mathcal{G}} \nonumber \\
   &+& \sum_I \grada b_I(\be,\bR_I) {\big(\bSt \be - \bSt \bR_I\big)\mkern-4mu}_\beta \hat{\phi}(\be,\bR) \bigg) \, \mathrm{d \be} \nonumber \\
   &=& \int_\Omega \bigg(\frac{1}{4 \pi} \grada \hat{\phi}(\be,\bR) \gradb \hat{\phi}(\be,\bR) + \sum_I \grada b_I(\be,\bR_I) {\big(\bSt \be - \bSt \bR_I\big)\mkern-4mu}_\beta \hat{\phi}(\be,\bR) \bigg) \, \mathrm{d \be} \nonumber \,,\\
\sigma \mkern-2mu ^{E_{self}}_{\alpha\beta} &=& \frac{\partial}{\partial \Fab} \Bigg( \frac{1}{2} \sum_I \int_\Omega b_I(\mathbf{Q} \be,\mathbf{Q} \bR_I) V_I (\mathbf{Q} \be,\mathbf{Q} \bR_I) \det(\bF) \, \mathrm{d \be} \Bigg) \Bigg|_{\mathrm{G}} \nonumber \\
&=& \frac{1}{2}\sum_I \int_\Omega \bigg(\grada b_I(\be,\bR_I){\big(\bSt \be - \bSt \bR_I\big)\mkern-4mu}_\beta V_I(\be,\bR_I) + b_I(\be,\bR_I) \grada V_I(\be,\bR_I){\big(\bSt \be - \bSt \bR_I\big)\mkern-4mu}_\beta \bigg)\, \mathrm{d \be}\nonumber \\
&+&  \delta_{\alpha \beta} E_{self}(\bR)  \nonumber \,.
\end{eqnarray}
The expression for $\sigma \mkern-2mu ^{E_{c}}_{\alpha\beta}$ can be found in Appendix~\ref{subsection:electrostaticcorrection}. In obtaining the terms $D_1$, $D_2$, and $D_3$, we have used the relation in Eqn.~\ref{Eqn:DerRhoFab}.  It is important to note that using Gauss' divergence theorem and the chain rule, it is possible to show that $\sigma \mkern-2mu ^{E_{self}}_{\alpha\beta} = 0$, consistent with the result obtained in the context of the planewave method \cite{Focher1994}. However, due to the inexact nature of the chain rule within the finite-difference approximation, $\sigma \mkern-2mu ^{E_{self}}_{\alpha\beta}$ can take significant values, as shown in Appendix~\ref{subsection:stresswithoutselfenergycontribution}. Therefore, we utilize the above formulation for $\sigma \mkern-2mu ^{E_{self}}_{\alpha\beta}$, which we have found to be particularly well suited for error cancellation with the other terms.  
%%%%%%%%%%%%%%%%%%%%%%%%%%%%%%%%%%%%%%%%%%%%%%%%%%%%%%%%%%%%%%%%%%%%%%%%%%%%%%%%%%%%%%%%%%%%%%%%%%%%%%%%%%%%%%%%%%%%%%%%%%%%%%%%%%%%%%%%%%%%%%%%%%%%%%%%%%%%%%%%%%%%%%%%%%%%%%%%
\subsection{Stress tensor contribution $\sigma \mkern-1mu ^{S}$}\label{Subsection:entropicstress}
The contribution to the stress tensor arising from the electronic entropy energy:
\begin{eqnarray}
\sigma \mkern-1mu ^{S}_{\alpha \beta} &=& \frac{\partial S^F(\bg)}{\partial \Fab} \Bigg|_{\mathcal{G}} \nonumber \\
&=& \frac{\partial}{\partial \Fab} \Bigg(\mkern-6mu - \, 2 \sigma  \sum_{n=1}^{N_s} \fint_{BZ^F} \Big(\occ_n(\bk^{F}) \log \occ_n(\bk^{F}) + \big(1 - \occ_n(\bk^{F})\big) \log \big(1 - \occ_n(\bk^{F}) \big) \Big) \, \mathrm{d \bk^{F}} \Bigg) \Bigg|_{\mathcal{G}} \nonumber \\
&=& \frac{\partial}{\partial \Fab} \Bigg(\mkern-6mu - 2 \sigma  \sum_{n=1}^{N_s} \fint_{BZ} \Big(\occ_n(\mathbf{Q} ^{\rm -T} \bk) \log \occ_n(\mathbf{Q} ^{\rm -T} \bk) + \big(1 - \occ_n(\mathbf{Q} ^{\rm -T} \bk)\big) \log \big(1 - \occ_n(\mathbf{Q} ^{\rm -T} \bk) \big) \Big) \, \mathrm{d \bk} \Bigg) \Bigg|_{\mathcal{G}} \nonumber \\
&=&  - 2 \sigma \sum_{n=1}^{N_s} \fint_{BZ} \frac{\partial \occ_n(\mathbf{Q} ^{\rm -T} \bk)}{\partial\Fab} \Bigg|_{\mathcal{G}} \log \Bigg( \frac{\hat{\occ}_n(\bk)}{1 - \hat{\occ}_n(\bk)} \Bigg) \, \mathrm{d \bk} \,. \label{Eqn:sigmaEent}
\end{eqnarray}
%%%%%%%%%%%%%%%%%%%%%%%%%%%%%%%%%%%%%%%%%%%%%%%%%%%%%%%%%%%%%%%%%%%%%%%%%%%%%%%%%%%%%%%%%%%%%%%%%%%%%%%%%%%%%%%%%%%%%%%%%%%%%%%%%%%%%%%%%%%%%%%%%%%%%%%%%%%%%%%%%%%%%%%%%%%%%%%%
\subsection{Stress tensor contribution $\sigma \mkern-2mu ^{\lambda_{mn}}$}\label{Subsection:orthonormalityconstraintstress}
The contribution  to the stress tensor arising from the constraint on the orthonormality of the orbitals:
\begin{eqnarray}
\sigma \mkern-2mu ^{\lambda_{mn}}_{\alpha \beta} &=& \frac{\partial}{\partial \Fab} \Bigg(2 \sum_{mn} \fint_{BZ^F} \lambda_{mn}(\bk^{F}) \bigg( \int_{\Omega^F} \psi_m^*(\be^{F},\bk^{F}) \psi_n(\be^{F},\bk^{F}) \, \mathrm{d \be^{F}} - \delta _{m n} \bigg) \,\mathrm{d \bk^{F}} \Bigg) \Bigg|_{\mathcal{G}} \nonumber \\
&=& \frac{\partial}{\partial\Fab} \Bigg(2 \sum_{mn} \fint_{BZ} \lambda_{mn}(\mathbf{Q} ^{\rm -T} \bk) \bigg( \int_{\Omega} \psi_m^*(\mathbf{Q} \be,\mathbf{Q} ^{\rm -T} \bk) \psi_n(\mathbf{Q} \be,\mathbf{Q} ^{\rm -T} \bk) \det(\bF) \, \mathrm{d \be} - \delta _{m n} \bigg) \,\mathrm{d \bk} \Bigg) \Bigg|_{\mathcal{G}} \nonumber \\
&=& F_1 + F_2 + F_3 + F_4 \,, \label{Eqn:sigmalambdamn}
\end{eqnarray}
where
\begin{eqnarray}
F_1 &=& 2 \sum_{mn} \fint_{BZ} \frac{\partial \lambda_{mn}(\mathbf{Q} ^{\rm -T} \bk)}{\partial \Fab} \Bigg|_{\mathcal{G}} \bigg(\int_{\Omega} \hat{\psi}_m^*(\be,\bk) \hat{\psi}_n(\be,\bk) \, \mathrm{d\be} - \delta _{m n} \bigg) \,\mathrm{d \bk} \nonumber \\
    &=& 0 \,, \nonumber \\
F_2 &=& 2 \sum_{m} \fint_{BZ} \hat{\lambda}_{m}(\bk) \int_{\Omega} \frac{\partial \psi_m^*(\mathbf{Q} \be,\mathbf{Q} ^{\rm -T} \bk)}{\partial \Fab} \bigg|_{\mathcal{G}}   \hat{\psi}_m(\be,\bk) \, \mathrm{d\be}  \, \mathrm{d \bk} \,, \nonumber \\
F_3 &=& 2 \sum_{n} \fint_{BZ} \hat{\lambda}_{n}(\bk) \int_{\Omega} \hat{\psi}_n^*(\be,\bk) \frac{\partial {\psi}_n(\mathbf{Q} \be,\mathbf{Q} ^{\rm -T} \bk)}{\partial \Fab} \bigg|_{\mathcal{G}} \, \mathrm{d\be} \,\mathrm{d \bk} \,, \nonumber \\
F_4 &=& 2 \sum_{m} \fint_{BZ} \hat{\lambda}_{m}(\bk) \int_{\Omega} \hat{\psi}_m^*(\be,\bk) \hat{\psi}_m(\be,\bk) \frac{\partial \big(\det(\bF)\big)}{\partial \Fab} \bigg|_{\mathcal{G}} \, \mathrm{d\be} \,\mathrm{d \bk} \,, \nonumber \\
    &=& 2 \sum_{m} \fint_{BZ} \hat{\lambda}_{m}(\bk) \int_{\Omega} \hat{\psi}_m^*(\be,\bk) \hat{\psi}_m(\be,\bk) \delta_{\alpha \beta} \, \mathrm{d\be} \,\mathrm{d \bk} \,. \nonumber
\end{eqnarray}
%%%%%%%%%%%%%%%%%%%%%%%%%%%%%%%%%%%%%%%%%%%%%%%%%%%%%%%%%%%%%%%%%%%%%%%%%%%%%%%%%%%%%%%%%%%%%%%%%%%%%%%%%%%%%%%%%%%%%%%%%%%%%%%%%%%%%%%%%%%%%%%%%%%%%%%%%%%%%%%%%%%%%%%%%%%%%%%%
\subsection{Stress tensor contribution $\sigma \mkern-2mu ^{\lambda_{f}}$}\label{Subsection:chargeconservationconstraintstress}
The contribution to the stress tensor arising from the constraint on the total number of electrons:
\begin{eqnarray}
\sigma \mkern-2mu ^{\lambda_{f}}_{\alpha \beta} &=& \frac{\partial}{\partial \Fab} \Bigg(\lambda_f \bigg( 2 \sum_{n=1}^{N_s} \fint_{BZ^F} \occ_n(\bk^{F}) \, \mathrm{d \bk} - N_e \bigg) \Bigg) \Bigg|_{\mathcal{G}} \nonumber \\
 &=& \frac{\partial}{\partial \Fab} \Bigg(\lambda_f \bigg( 2 \sum_{n=1}^{N_s} \fint_{BZ} \occ_n(\mathbf{Q} ^{\rm -T} \bk) \, \mathrm{d \bk} - N_e \bigg) \Bigg) \Bigg|_{\mathcal{G}} \nonumber \\
&=& 2\,\hat{\lambda}_f \sum_{n=1}^{N_s} \fint_{BZ} \frac{\partial \occ_n(\mathbf{Q} ^{\rm -T} \bk)}{\partial\Fab} \Bigg|_{\mathcal{G}} \, \mathrm{d \bk} \,. \label{Eqn:sigmalambdaf}
\end{eqnarray}
%%%%%%%%%%%%%%%%%%%%%%%%%%%%%%%%%%%%%%%%%%%%%%%%%%%%%%%%%%%%%%%%%%%%%%%%%%%%%%%%%%%%%%%%%%%%%%%%%%%%%%%%%%%%%%%%%%%%%%%%%%%%%%%%%%%%%%%%%%%%%%%%%%%%%%%%%%%%%%%%%%%%%%%%%%%%%%%%
\subsection{Total stress tensor}\label{Subsection:completestresstensor}
It follows from Eqns.~\ref{Eqn:stressdefinition} and~\ref{Eqn:lagrangian} that the total stress can be written in terms of the various contributions derived in the previous subsections as 
\begin{eqnarray}
\sab &=& \frac{1}{|\Omega|} \bigg[\sigma \mkern-2mu ^{T_s}_{\alpha \beta} + \sigma \mkern-2mu ^{E_{xc}}_{\alpha \beta} + \sigma \mkern-2mu ^{E_{nl}}_{\alpha\beta} + \sigma \mkern-2mu ^{E_{el}}_{\alpha \beta} - \sigma \mkern-1.5mu ^{S}_{\alpha \beta} - \sigma \mkern-1.5mu ^{\lambda_{mn}}_{\alpha \beta} - \sigma \mkern-2mu ^{\lambda_{f}}_{\alpha \beta} \bigg]  \nonumber \\
&=& \frac{1}{|\Omega|} \Bigg[ \sum_{i=1}^5 (A_i + B_i + C_i + D_i) - \sigma \mkern-2mu ^{E_{self}}_{\alpha\beta} + \sigma \mkern-2mu ^{E_{c}}_{\alpha\beta} - \sigma \mkern-1.5mu ^{S}_{\alpha \beta} - \sum_{i=1}^4 F_i - \sigma \mkern-2mu ^{\lambda_{f}}_{\alpha \beta}  \Bigg] \,. \label{Eqn:stressexpansion}
\end{eqnarray}
As a consequence of the Euler-Lagrange equations in Eqns.~\ref{Eqn:eulerlagrange}--\ref{Eqn:poissonequation}, it follows that
\begin{eqnarray}
A_1 + B_1 + C_1 + D_1 - \sigma \mkern-1.5mu ^{S}_{\alpha \beta} - \sigma \mkern-2mu ^{\lambda_{f}}_{\alpha \beta} & = & 0 \,, \\
A_2 + B_2 + C_2 + D_2 - F_2 &=& 0 \, \\
A_3 + B_3 + C_3 + D_3 - F_3 &=& 0 \,, \\
A_4 + B_4 + C_4 + D_4 - F_4 &=& \delta_{\alpha \beta} \, \bigg[ E_{xc}(\hat{\rho}, \bm{\nabla} \hat{\rho}) - \int_\Omega V_{xc} \big(\hat{\rho}(\be),\bm{\nabla} \hat{\rho}(\be)\big) \hat{\rho}(\be) \, \mathrm{d \be} + E_{nl}(\hat{\bPsi},\hat{\bg},\bR) \nonumber \\
&+& \frac{1}{2} \int_\Omega \big(b(\be,\bR) - \hat{\rho}(\be)\big) \hat{\phi}(\be,\bR) \, \mathrm{d \be} \bigg] \,.
\end{eqnarray}
Inserting the above relations and those from Eqns.~\ref{Eqn:sigmaTs},~\ref{Eqn:sigmaExc},~\ref{Eqn:sigmaEnl},~\ref{Eqn:sigmaEel},~\ref{Eqn:sigmaEent},~\ref{Eqn:sigmalambdamn}, and~\ref{Eqn:sigmalambdaf} into Eqn.~\ref{Eqn:stressexpansion}, we arrive at the expression for the total stress: 
\begin{eqnarray}
\sigma \mkern-2mu_{\alpha \beta} &=& \frac{1}{|\Omega |} \Bigg[ -2 \sum_{n=1}^{N_s} \fint_{BZ} \hat{\occ}_n(\bk) \int_\Omega \grada \hat{\psi}_n^*(\be,\bk) \gradb \hat{\psi}_n(\be,\bk) \, \mathrm{d \be} \, \mathrm{d \bk} + \delta_{\alpha \beta} \bigg( E_{xc}(\hat{\rho},\bm{\nabla} \hat{\rho}) - \int_\Omega V_{xc}\big(\hat{\rho}(\be),\bm{\nabla} \hat{\rho}(\be)\big) \hat{\rho}(\be) \, \mathrm{d \be} \bigg) \nonumber \\
&-& \int_\Omega \hat{\rho}(\be) \frac{\partial \varepsilon_{xc}\big(\hat{\rho}(\be),\bm{\nabla} \hat{\rho}(\be) \big)}{\partial \big(\gradb \hat{\rho}(\be)\big)} \grada \hat{\rho}(\be) \, \mathrm{d \be} -\delta_{\alpha \beta} E_{nl}(\hat{\bPsi},\hat{\bg},\bR) - 4 \sum_{n=1}^{N_s} \fint_{BZ} \hat{\occ}_n(\bk) \sum_J \sum_{lm} \gammaJl \Re \Bigg[ \Bigg(\sum_{J'} \int_\Omega \pchis(\be,\bRJp) \nonumber \\
&\times & e^{i \bk \cdot (\bRJ - \bRJp)} {\big(\bSt \be - \bSt \bRJp \big)\mkern-4mu}_\beta \grada \hat{\psi}_n(\be,\bk)  \, \mathrm{d \be} \Bigg) \Bigg(\int_\Omega \tchins(\be,\bRJ,\bk) \hat{\psi}^*_n(\be,\bk) \, \mathrm{d \be}\Bigg) \Bigg] \, \mathrm{d \bk} + \frac{1}{4 \pi} \int_\Omega \grada \hat{\phi}(\be,\bR) \gradb \hat{\phi}(\be,\bR) \, \mathrm{d \be} \nonumber \\
&+& \sum_I \int_\Omega \grada b_I(\be,\bR_I) {\big(\bSt \be - \bSt \bR_I \big)\mkern-4mu}_\beta \Big(\hat{\phi}(\be,\bR) - \frac{1}{2} \, V_I(\be,\bR_I) \Big) \, \mathrm{d \be} -\frac{1}{2} \sum_I \int_\Omega \grada V_I(\be,\bR_I) {\big(\bSt \be - \bSt \bR_I \big)\mkern-4mu}_\beta b_I(\be,\bR_I) \, \mathrm{d \be} \nonumber \\
&+& \frac{1}{2} \delta_{\alpha \beta} \int_\Omega \big(b(\be,\bR) - \hat{\rho}(\be) \big) \hat{\phi}(\be,\bR) \, \mathrm{d \be} -\delta_{\alpha \beta} E_{self}(\bR) + \sigma \mkern-2mu ^{E_c}_{\alpha\beta} \Bigg] \,.
\end{eqnarray}
The pressure can therefore be written as: 
\begin{eqnarray}
P &=& -\frac{1}{3} \sum_{i=1}^3 \sigma \mkern-2mu _{ii} \nonumber \\
&=& -\frac{1}{3 |\Omega |} \Bigg[ -4 \sum_{n=1}^{N_s} \fint_{BZ} \hat{\occ}_n(\bk) \hat{\lambda}_n(\bk) \, \mathrm{d \bk} + 3 E_{xc}(\hat{\rho},\bm{\nabla} \hat{\rho}) - \int_\Omega \bigg( V_{xc}\big(\hat{\rho}(\be),\bm{\nabla} \hat{\rho}(\be)\big) + \frac{\partial \varepsilon_{xc}\big(\hat{\rho}(\be),\bm{\nabla} \hat{\rho}(\be) \big)}{\partial \big(\bm{\nabla} \hat{\rho}(\be)\big)} \cdot \bm{\nabla} \hat{\rho}(\be) \bigg) \hat{\rho}(\be)  \, \mathrm{d \be} \nonumber \\
&-& E_{nl}(\hat{\bPsi},\hat{\bg},\bR) - 4 \sum_{n=1}^{N_s} \fint_{BZ} \hat{\occ}_n(\bk) \sum_J \sum_{lm} \gammaJl \Re \Bigg[ \Bigg(\sum_{J'} \int_\Omega \pchis(\be,\bRJp) e^{i \bk \cdot (\bRJ - \bRJp)} \big(\bSt \be - \bSt \bRJp \big) \cdot \bm{\nabla} \hat{\psi}_n(\be,\bk)  \, \mathrm{d \be} \Bigg) \nonumber \\
&\times & \Bigg(\int_\Omega \tchins(\be,\bRJ,\bk) \hat{\psi}^*_n(\be,\bk) \, \mathrm{d \be}\Bigg) \Bigg] \, \mathrm{d \bk} + \frac{1}{4 \pi} \int_\Omega \big| \bm{\nabla} \hat{\phi}(\be,\bR) \big|^2 \, \mathrm{d \be} + \sum_I \int_\Omega \bm{\nabla} b_I(\be,\bR_I) \cdot \big(\bSt \be - \bSt \bR_I\big) \nonumber \\
& \times & \Big(\hat{\phi}(\be,\bR) - \frac{1}{2} V_I(\be,\bR_I) \Big) \, \mathrm{d \be} - \frac{1}{2} \sum_I \int_\Omega \bm{\nabla} V_I(\be,\bR_I) \cdot \big(\bSt \be - \bSt  \bR_I\big) b_I(\be,\bR_I) \, \mathrm{d \be} + \frac{1}{2} \int_\Omega \big(\hat{\rho}(\be) + 3 b(\be,\bR) \big) \hat{\phi}(\be,\bR) \, \mathrm{d \be} \nonumber \\
&-& 3 E_{self}(\bR) + \sum_{i=1}^3 \sigma \mkern-2mu ^{E_c}_{ii} \Bigg] \,.
\end{eqnarray} 
The above expressions for the stress and pressure can be evaluated in $\mathcal{O}(N)$ operations \footnote{The locality of the operations also makes them particularly well suited to scalable parallel computations.} and are applicable for the general case of a non-orthogonal crystal system with Brillouin zone sampling and for the choice of a semilocal exchange-correlation functional. Indeed, the expression for a $\Gamma$-point calculation can be obtained by dropping the volume-average integral over the Brillouin zone and setting $\bk = \mathbf{0}$ in the expressions. In addition, the expression for the choice of a local exchange-correlation functional such as the local density approximation (LDA) \cite{Kohn1965} can be obtained by dropping $ \bm{\nabla} \hat{\rho}$ and the associated derivative terms.  
   
%%%%%%%%%%%%%%%%%%%%%%%%%%%%%%%%%%%%%%%%%%%%%%%%%%%%%%%%%%%%%%%%%%%%%%%%%%%%%%%%%%%%%%%%%%%%%%%%%%%%%%%%%%%%%%%%%%%%%%%%%%%%%%%%%%%%%%%%%%%%%%%%%%%%%%%%%%%%%%%%%%%%%%%%%%%%%
%%%%%%%%%%%%%%%%%%%%%%%%%%%%%%%%%%%%%%%%%%%%%%%%%%%%%%%%%%%%%%%%%%%%%%%%%%%%%%%%%%%%%%%%%%%%%%%%%%%%%%%%%%%%%%%%%%%%%%%%%%%%%%%%%%%%%%%%%%%%%%%%%%%%%%%%%%%%%%%%%%%%%%%%%%%%%
%%%%%%%%%%%%%%%%%%%%%%%%%%%%%%%%%%%%%%%%%%%%%%%%%%%%%%%%%%%%%%%%%%%%%%%%%%%%%%%%%%%%%%%%%%%%%%%%%%%%%%%%%%%%%%%%%%%%%%%%%%%%%%%%%%%%%%%%%%%%%%%%%%%%%%%%%%%%%%%%%%%%%%%%%%%%%
\section{Examples and results}\label{Section:implementation}
In this section, we verify the accuracy and efficiency of the proposed formulation of the stress tensor for real-space DFT calculations. To do so, we incorporate it into the M-SPARC prototype code, a serial implementation of the large-scale real-space DFT code SPARC \cite{ghosh2016sparc1,ghosh2016sparc2}. The Poisson problem in Eqn.~\ref{Eqn:poissonequation} is solved using the Alternating Anderson-Richardson (AAR) method \cite{pratapa2016anderson,suryanarayana2016alternating}. The electronic ground-state is calculated using the Chebyshev-filtered subspace iteration (CheFSI) \cite{Zhou2006parallel,Zhou2006self}, with acceleration provided by the restarted Periodic Pulay method \cite{Pratapa2015restarted,banerjee2016periodic}. In all simulations, we employ a twelfth-order accurate finite-difference discretization, norm-conserving Troullier-Martins pseudopotentials \cite{Troullier}, trapezoidal rule for all integrations in real space, and the Monkhorst-Pack \cite{Monkhorst} grid for integration over the Brillouin zone. A more detailed description of the underlying finite-difference formulation and implementation can be found in our previous work \cite{ghosh2016sparc1,ghosh2016sparc2}.

As representative examples, we consider the following systems: (i) hexagonal close packed (hcp) titanium with equilibrium lattice parameters: $L_1 = 5.47$ Bohr, $L_2 = 5.47$ Bohr, $L_3 = 8.85$ Bohr, $\theta_1 = \theta_2 = 90^{\circ}$, and $\theta_3 = 120^{\circ}$; (ii) diamond cubic (dc) germanium with equilibrium lattice parameters: $L_1 = L_2 = L_3 = 10.74$ Bohr, and $\theta_1 = \theta_2 = \theta_3 = 90^{\circ}$; and (iii) triclinic titanium with equilibrium lattice parameters: $L_1 = 5.47$ Bohr, $L_2 = 5.47$ Bohr, $L_3 = 8.85$ Bohr, $\theta_1 = 97^{\circ}$, $\theta_2 = 82^{\circ}$, and $\theta_3 = 107^{\circ}$. We employ the PW \cite{Perdew1992accurate} variant of LDA and the PBE \cite{perdew1996generalized} variant of GGA as exchange-correlation functionals for the titanium and germanium systems, respectively \footnote{The exchange-correlation functionals are incorporated via an interface with the libxc \cite{libxc} package.}. Wherever suitable, we compare with the planewave code ABINIT \cite{ABINIT}, wherein we use planewave cutoffs of 70 Ha and 30 Ha for the titanium and germanium systems, respectively, resulting in stresses that are converged to within 0.01\%.   

%%%%%%%%%%%%%%%%%%%%%%%%%%%%%%%%%%%%%%%%%%%%%%%%%%%%%%%%%%%%%%%%%%%%%%%%%%%%%%%%%%%%%%%%%%%%%%%%%%%%%%%%%%%%%%%%%%%%%%%%%%%%%%%%%%%%%%%%%%%%%%

\subsection{Convergence of stress tensor with discretization}\label{Subsection:convergence}
\begin{figure}[h!]
\includegraphics[keepaspectratio=true,scale=0.5]{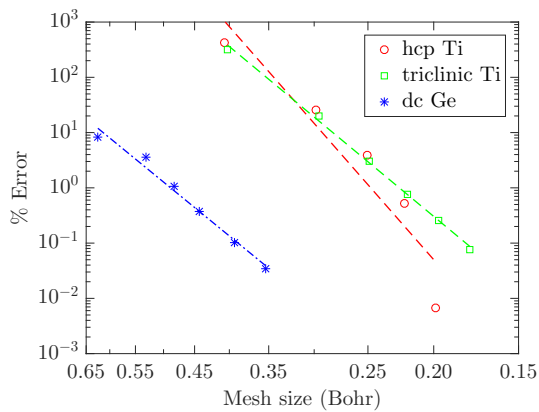}
\caption{\label{Fig:convergenceplot} Convergence of the stress tensor with mesh size for the hcp titanium, triclinic titanium, and dc germanium systems. The error is defined to be magnitude of the maximum difference in any component. The straight lines represent linear fits to the data.}
\end{figure}

First, we verify convergence of the stress tensor with respect to spatial discretization.  For this study, we choose a 2-atom unit cell of hcp titanium uniformly expanded by $1\%$, a 2-atom unit cell of triclinic titanium, and an 8-atom unit cell of dc germanium uniformly compressed by $1\%$. We employ $6 \times 6 \times 6$ and $5 \times 5 \times 5$ $\mathbf{k}$-point grids for the titanium and germanium systems, respectively. In Fig.~\ref{Fig:convergenceplot}, we present the error in the calculated stress tensor as a function of mesh size. The error is defined with respect to reference M-SPARC results that are converged to 0.001\% accuracy, which in turn match ABINIT results to within 0.2\%. It is clear that there is systematic convergence of the computed stress tensor. On performing a linear fit to the data, we obtain convergence rates of approximately $\mathcal{O} (h^{10}$) with respect to mesh size. These results demonstrate that high rates of convergence---similar to those obtained for the energy and atomic forces \cite{ghosh2016sparc2,AbhirajKP}---can be obtained for the stress tensor within the proposed formulation. 

%%%%%%%%%%%%%%%%%%%%%%%%%%%%%%%%%%%%%%%%%%%%%%%%%%%%%%%%%%%%%%%%%%%%%%%%%%%%%%%%%%%%%%%%%%%%%%%%%%%%%%%%%%%%%%%%%%%%%%%%%%%%%%%%%%%%%%%%%%%%%%

\subsection{Cell optimization using the stress tensor}\label{Subsection:celloptimization}
\begin{figure}[h!]
\subfloat[Computed energy difference and its cubic spline fit]{\includegraphics[keepaspectratio=true,scale=0.5]{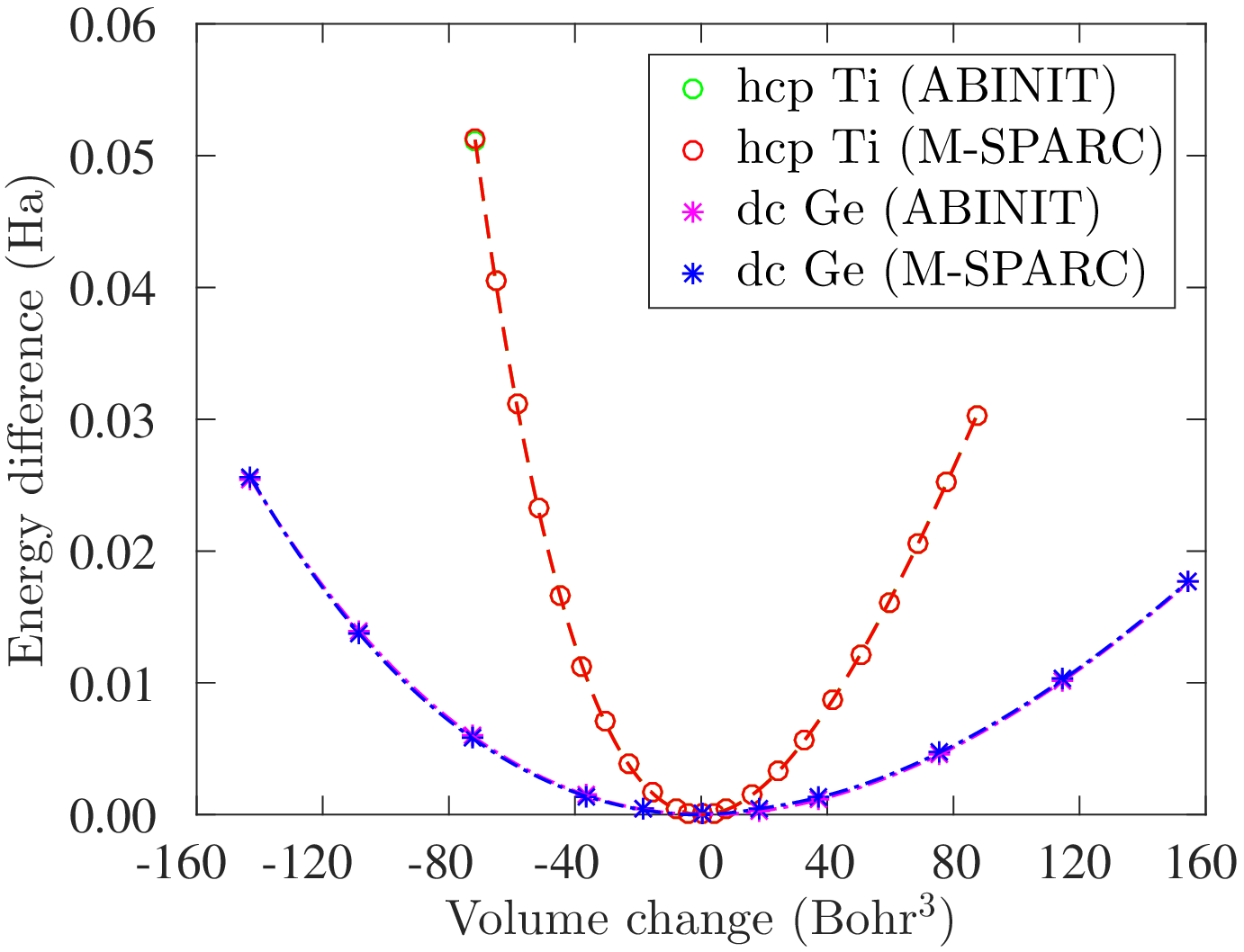} \label{Fig:consistencyplots1} } 
\subfloat[Computed pressure and the derivative of the cubic spline fit to the energy]{\includegraphics[keepaspectratio=true,scale=0.5]{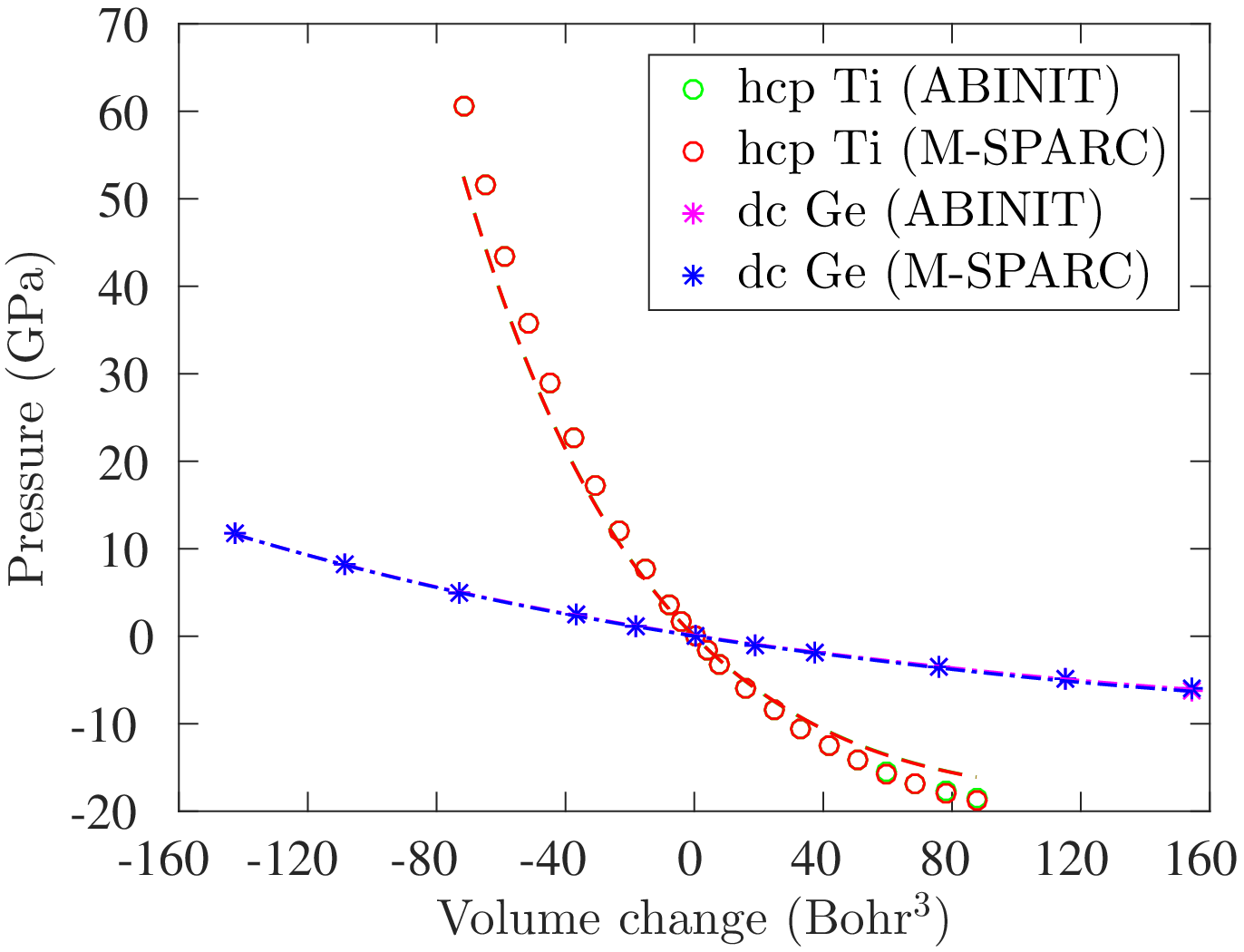} \label{Fig:consistencyplots2}}
\caption{\label{Fig:consistencyplots} Variation in the energy and pressure computed by ABINIT and M-SPARC as a function of volume for the hcp titanium and dc germanium systems. The volume change and energy difference are defined with respect to the equilibrium system, i.e., cell corresponding to zero stress.}
\end{figure}

Next, we verify the accuracy of the computed stress tensor for performing cell optimization. For this study, we consider a 2-atom unit cell of hcp titanium and an 8-atom unit cell of dc germanium, with $6 \times 6 \times 6$ and $5 \times 5 \times 5$ $\mathbf{k}$-point grids for Brillouin zone integration, respectively. In Fig.~\ref{Fig:consistencyplots}, we plot the variation in energy and pressure versus the unit cell volume as computed by M-SPARC and ABINIT. Specifically, we plot the computed energy and its cubic spline fit in Fig.~\ref{Fig:consistencyplots1}, and the computed pressure and the derivative of the cubic spline fit to the energy in Fig.~\ref{Fig:consistencyplots2}. Note that we have employed a constant number of grid points in M-SPARC, i.e., they are independent of the unit cell volume and correspond to mesh sizes of 0.22 and 0.44 Bohr for the equilibrium titanium and germanium systems, respectively. It is clear from the results that there is excellent agreement between ABINIT and M-SPARC, with the results being practically indistinguishable. In particular, as determined from the data in Fig.~\ref{Fig:consistencyplots2}, the difference in equilibrium lattice constants predicted  by M-SPARC and ABINIT for the titanium and germanium systems are 0.0003 Bohr and 0.003 Bohr, respectively, and the corresponding difference in the bulk modulus is 0.004 GPa and 0.3 GPa, respectively. The Pulay stress \cite{Pulay1969} at the chosen mesh sizes is estimated to be  0.014 GPa and 0.0086 GPa for the titanium and germanium systems, respectively.\footnote{The Pulay stress is estimated using the technique described in the VASP \cite{VASP} manual.}

It is also clear from the results in Fig.~\ref{Fig:consistencyplots} that the computed energy and pressure are consistent within the proposed formulation. This is also true for the complete stress tensor, as verified by the results in Table~\ref{Table:stressconsistency} for the 2-atom triclinic titanium system with $6 \times 6 \times 6$ $\mathbf{k}$-point sampling and mesh size of 0.22 Bohr. In particular, the maximum difference between the computed stress tensor and that obtained from the numerical derivative of the energy is less than 1\%. 

\begin{table}[h!]
\begin{tabular}{lccccccc}
\hline
& \quad \quad \quad & $\sigma_{11}$ \quad \quad \quad & $\sigma_{12}$  \quad \quad \quad & $\sigma_{13}$  \quad \quad \quad & $\sigma_{22}$  \quad \quad \quad & $\sigma_{23}$  \quad \quad \quad & $\sigma_{33}$  \\
\hline
& Computed \quad \quad \quad & $6.884$ \quad \quad \quad & $4.371$  \quad \quad \quad & $-3.237$  \quad \quad \quad & $9.610$  \quad \quad \quad & $2.476$  \quad \quad \quad & $5.104$ \\
& Numerical derivative \quad \quad \quad & $ 6.852$ \quad \quad \quad & $4.351$  \quad \quad \quad & $-3.260$  \quad \quad \quad & $9.551$  \quad \quad \quad & $2.493$  \quad \quad \quad & $5.053$ \\
\hline
\end{tabular}
\caption{\label{Table:stressconsistency} Computed stress tensor and that obtained from the numerical derivative of the energy for the triclinic titanium system by M-SPARC. All stress component values are reported in GPa.}
\end{table}

%%%%%%%%%%%%%%%%%%%%%%%%%%%%%%%%%%%%%%%%%%%%%%%%%%%%%%%%%%%%%%%%%%%%%%%%%%%%%%%%%%%%%%%%%%%%%%%%%%%%%%%%%%%%%%%%%%%%%%%%%%%%%%%%%%%%%%%%%%%%%
\subsection{Stress tensor in ab-initio molecular dynamics}\label{Subsection:largesystem}
Finally, we verify the ability of the proposed formulation to accurately calculate the stress tensor in AIMD simulations. To do so, we consider $128$-atom hcp titanium and $216$-atom dc germanium systems with the atoms randomly perturbed by up to $10 \%$ of nearest neighbor distance and perform $\Gamma$-point calculations, as is typical in AIMD simulations. In M-SPARC, we employ mesh sizes of  $h= 0.21$ Bohr and $h= 0.44$ Bohr for the titanium and germanium systems, respectively. It is clear from the results presented in Table~\ref{Table:stresslargesystem} that there is very good agreement between M-SPARC and ABINIT, with the maximum difference in any stress component being 0.9 \%, an accuracy representative of those desired in practical calculations.  Note that as the mesh is refined in M-SPARC, the agreement with ABINIT further increases. Also note that the calculation of the stress tensor takes less than $1 \%$ of the total simulation time in M-SPARC, which verifies the efficiency of the proposed  formulation for real-space DFT calculations.
\begin{table}[h!]
\begin{tabular}{llccccccc}
\hline
& \quad \quad \quad & \quad \quad \quad & $\sigma_{11}$ \quad \quad \quad & $\sigma_{12}$  \quad \quad \quad & $\sigma_{13}$  \quad \quad \quad & $\sigma_{22}$  \quad \quad \quad & $\sigma_{23}$  \quad \quad \quad & $\sigma_{33}$  \\
\hline
\multirow{2}{*}{Ti$_{128}$} &\quad \quad \quad & M-SPARC \quad \quad \quad & $-6.175$ \quad \quad \quad & $0.585$  \quad \quad \quad & $0.235$  \quad \quad \quad & $-5.219$  \quad \quad \quad & $0.000$  \quad \quad \quad & $-5.981$ \\
~& \quad \quad \quad & ABINIT \quad \quad \quad & $-6.159$ \quad \quad \quad & $0.580$  \quad \quad \quad & $0.235$  \quad \quad \quad & $-5.251$  \quad \quad \quad & $0.000$  \quad \quad \quad & $-6.038$ \\
\hline
\multirow{2}{*}{Ge$_{216}$} &\quad \quad & M-SPARC \quad \quad \quad & $-23.569$ \quad \quad \quad & $1.933$  \quad \quad \quad & $2.551$  \quad \quad \quad & $-27.397$  \quad \quad \quad & $-3.746$  \quad \quad \quad & $-25.032$  \\
~&\quad \quad \quad & ABINIT \quad \quad \quad & $-23.569$ \quad \quad \quad & $1.933$  \quad \quad \quad & $2.551$  \quad \quad \quad & $-27.398$  \quad \quad \quad & $-3.746$  \quad \quad \quad & $-25.034$  \\
\hline
\end{tabular}
\caption{\label{Table:stresslargesystem} Stress tensor computed by M-SPARC and ABINIT for the hcp titanium and dc germanium systems. All stress component values are reported in GPa.}
\end{table}
%%%%%%%%%%%%%%%%%%%%%%%%%%%%%%%%%%%%%%%%%%%%%%%%%%%%%%%%%%%%%%%%%%%%%%%%%%%%%%%%%%%%%%%%%%%%%%%%%%%%%%%%%%%%%%%%%%%%%%%%%%%%%%%%%%%%%%%%%%%%%%%%%%%%%%%%%%%%%%%%%%%%%%%%%%%%%
%%%%%%%%%%%%%%%%%%%%%%%%%%%%%%%%%%%%%%%%%%%%%%%%%%%%%%%%%%%%%%%%%%%%%%%%%%%%%%%%%%%%%%%%%%%%%%%%%%%%%%%%%%%%%%%%%%%%%%%%%%%%%%%%%%%%%%%%%%%%%%%%%%%%%%%%%%%%%%%%%%%%%%%%%%%%%
%%%%%%%%%%%%%%%%%%%%%%%%%%%%%%%%%%%%%%%%%%%%%%%%%%%%%%%%%%%%%%%%%%%%%%%%%%%%%%%%%%%%%%%%%%%%%%%%%%%%%%%%%%%%%%%%%%%%%%%%%%%%%%%%%%%%%%%%%%%%%%%%%%%%%%%%%%%%%%%%%%%%%%%%%%%%%

\section{Concluding remarks}\label{Section:conclusion}
In this work, we have presented an accurate and efficient formulation of the stress tensor for Kohn-Sham DFT calculations employing the real-space finite-difference method. Specifically, while making use of a local formulation of the electrostatics, we have derived a linear-scaling expression for the stress tensor that is applicable to simulations with unit cells of arbitrary symmetry, semilocal exchange-correlation functionals, and Brillouin zone integration. In particular, we have rewritten the contributions to the stress tensor arising from the self energy and the nonlocal pseudopotential energy so as to make them amenable to the real-space method, thereby achieving up to three orders of magnitude improvement in the accuracy of the computed stresses. Through selected examples that are representative of static and dynamic DFT calculations, we have verified the accuracy and efficiency of the derived expression. In particular, we have demonstrated that the proposed formulation  obtains high rates of convergence with spatial discretization and that there is consistency between the computed energy and stress tensor, while maintaining very good agreement with reference planewave results. Overall, this paper overcomes one of the limitations of real-space approaches, i.e., inability to compute the stress tensor, making them an even more attractive choice for DFT calculations.

%%%%%%%%%%%%%%%%%%%%%%%%%%%%%%%%%%%%%%%%%%%%%%%%%%%%%%%%%%%%%%%%%%%%%%%%%%%%%%%%%%%%%%%%%%%%%%%%%%%%%%%%%%%%%%%%%%%%%%%%%%%%%%%%%%%%%%%%%%%%%%%%%%%%%%%%%%%%%%%%%%%%%%%%%%%%%
%%%%%%%%%%%%%%%%%%%%%%%%%%%%%%%%%%%%%%%%%%%%%%%%%%%%%%%%%%%%%%%%%%%%%%%%%%%%%%%%%%%%%%%%%%%%%%%%%%%%%%%%%%%%%%%%%%%%%%%%%%%%%%%%%%%%%%%%%%%%%%%%%%%%%%%%%%%%%%%%%%%%%%%%%%%%%
%%%%%%%%%%%%%%%%%%%%%%%%%%%%%%%%%%%%%%%%%%%%%%%%%%%%%%%%%%%%%%%%%%%%%%%%%%%%%%%%%%%%%%%%%%%%%%%%%%%%%%%%%%%%%%%%%%%%%%%%%%%%%%%%%%%%%%%%%%%%%%%%%%%%%%%%%%%%%%%%%%%%%%%%%%%%%
\section*{ACKNOWLEDGEMENTS}
The authors gratefully acknowledge the support of the National Science Foundation (CAREER - 1553212). The authors also acknowledge John E. Pask for helpful discussions and for bringing some important references to our attention.  

%%%%%%%%%%%%%%%%%%%%%%%%%%%%%%%%%%%%%%%%%%%%%%%%%%%%%%%%%%%%%%%%%%%%%%%%%%%%%%%%%%%%%%%%%%%%%%%%%%%%%%%%%%%%%%%%%%%%%%%%%%%%%%%%%%%%%%%%%%%%%%%%%%%%%%%%%%%%%%%%%%%%%%%%%%%%%
%%%%%%%%%%%%%%%%%%%%%%%%%%%%%%%%%%%%%%%%%%%%%%%%%%%%%%%%%%%%%%%%%%%%%%%%%%%%%%%%%%%%%%%%%%%%%%%%%%%%%%%%%%%%%%%%%%%%%%%%%%%%%%%%%%%%%%%%%%%%%%%%%%%%%%%%%%%%%%%%%%%%%%%%%%%%%
%%%%%%%%%%%%%%%%%%%%%%%%%%%%%%%%%%%%%%%%%%%%%%%%%%%%%%%%%%%%%%%%%%%%%%%%%%%%%%%%%%%%%%%%%%%%%%%%%%%%%%%%%%%%%%%%%%%%%%%%%%%%%%%%%%%%%%%%%%%%%%%%%%%%%%%%%%%%%%%%%%%%%%%%%%%%%

\appendix

\section{Stress tensor contribution $\sigma \mkern-1mu ^{E_c}$}\label{subsection:electrostaticcorrection}
The repulsive energy correction for overlapping pseudocharges takes the form \cite{Suryanarayana2014524,ghosh2016higher}:
\begin{equation}\label{electrostaticcorrectionenergy}
E_{c}(\bR) = \frac{1}{2} \int_\Omega \big(\tilde{b}(\be,\bR) + b(\be,\bR) \big) V_c(\be,\bR) \, \mathrm{d \be} + \frac{1}{2} \sum_I \int_\Omega b_I(\be,\bR_I) V_I(\be,\bR_I) \, \mathrm{d \be} - \frac{1}{2} \sum_I \int_\Omega \tilde{b}_I(\be,\bR_I) \tilde{V}_I(\be,\bR_I) \, \mathrm{d \be} \,, \\ 
\end{equation}
where $V_c(\be,\bR) = \sum_I \big( \tilde{V_I}(\be,\bR_I) - V_I(\be,\bR_I) \big)$; $\tilde{b} = \sum_I \tilde{b}_I$ denotes the total reference pseudocharge density of the nuclei with $\tilde{b}_I$ being the reference pseudocharge density of the $I^{th}$ nucleus that generates the potential $\tilde{V}_I$; and the summation index $I$ runs over all atoms in $\R^3$. The contribution to the stress tensor arising from this repulsive energy correction: 
\begin{eqnarray}
\sigma \mkern-2mu ^{E_{c}}_{\alpha\beta} &=& \frac{\partial E_c(\bR \mkern-2mu^{F})}{\partial \Fab} \bigg|_{\mathcal{G}}\nonumber \\
&=& \frac{\partial}{\partial \Fab} \Bigg( \frac{1}{2} \int_{\Omega^F} \big( \tilde{b}(\be^F, \bR \mkern-2mu^F) + b(\be^F,\bR\mkern-2mu ^F) \big) V_c(\be^F,\bR\mkern-2mu ^F) \, \mathrm{d \be^F} + \frac{1}{2} \sum_I \int_{\Omega^F} b_I(\be^F,\bR_I^F) V_I(\be^F,\bR_I^F) \, \mathrm{d \be^F} \nonumber \\
&-& \frac{1}{2} \sum_I \int_{\Omega^F} \tilde{b}_I(\be^F,\bR_I^F) \tilde{V}_I(\be^F,\bR_I^F) \, \mathrm{d \be^F} \Bigg) \Bigg|_{\mathcal{G}} \nonumber \\
&=& \frac{\partial}{\partial \Fab} \Bigg( \frac{1}{2} \int_\Omega \big( \tilde{b}(\mathbf{Q} \be,\mathbf{Q} \bR) + b(\mathbf{Q} \be,\mathbf{Q} \bR) \big) V_c(\mathbf{Q} \be,\mathbf{Q} \bR) \det(\bF) \, \mathrm{d \be} + \frac{1}{2} \sum_I \int_\Omega b_I(\mathbf{Q} \be,\mathbf{Q} \bR_I) V_I(\mathbf{Q} \be,\mathbf{Q} \bR_I) \det(\bF) \, \mathrm{d \be} \nonumber \\
&-& \frac{1}{2} \sum_I \int_\Omega \tilde{b}_I(\mathbf{Q} \be,\mathbf{Q} \bR_I) \tilde{V}_I(\mathbf{Q} \be,\mathbf{Q} \bR_I) \det(\bF) \, \mathrm{d \be} \Bigg) \Bigg|_{\mathcal{G}} \nonumber \\
&=& \frac{1}{2} \sum_I \int_\Omega \bigg(\grada \tilde{b}_I(\be,\bR_I) \big(V_c(\be,\bR) - \tilde{V}_I(\be,\bR_I) \big) + \grada b_I(\be,\bR_I) \big(V_c(\be,\bR) + V_I(\be,\bR_I) \big) +  \big(b(\be,\bR) + \tilde{b}(\be,\bR) \big) \nonumber \\
&\times & \big( \grada \tilde{V}_I(\be,\bR_I) - \grada V_I(\be,\bR_I) \big) -\grada \tilde{V}_I(\be,\bR_I) \tilde{b}_I(\be,\bR_I) + \grada V_I(\be,\bR_I) b_I(\be,\bR_I) \bigg) {\big(\bSt \be - \bSt \bR_I \big)\mkern-4mu}_\beta \, \mathrm{d \be} \nonumber \\
&+& \delta_{\alpha \beta} E_c(\bR) \,.
\end{eqnarray} 
As discussed in Appendix~\ref{subsection:stresswithoutselfenergycontribution}, the contribution to the stress tensor arising from the self energy terms are not identically zero within the finite-difference approximation, and therefore have been retained to ensure the accuracy of the proposed formulation. 

%%%%%%%%%%%%%%%%%%%%%%%%%%%%%%%%%%%%%%%%%%%%%%%%%%%%%%%%%%%%%%%%%%%%%%%%%%%%%%%%%%%%%%%%%%%%%%%%%%%%%%%%%%%%%%%%%%%%%%%%%%%%%%%%%%%%%%%%%%%%%%%
\section{On the reformulation of the stress tensor contribution $\sigma \mkern-2mu ^{E_{nl}}$ }\label{subsection:stresswithoutnonlocalreformulation}

\begin{figure}[h!]
\includegraphics[keepaspectratio=true,scale=0.5]{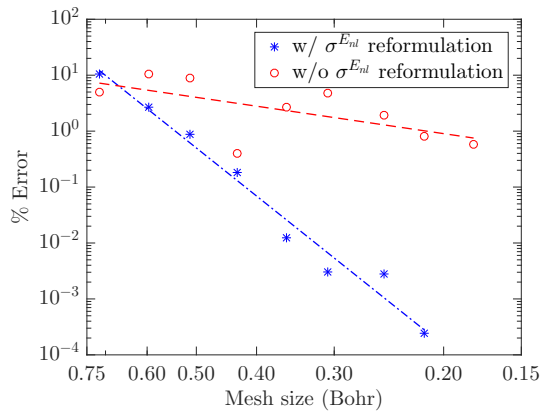}
\caption{\label{Fig:NLPcomparison} Convergence of the stress tensor with mesh size for the dc germanium system with and without the nonlocal reformulation. The error is defined to be magnitude of the maximum difference in any component. The straight lines represent linear fits to the data.}
\end{figure}

In Section~\ref{Subsection:nonlocalpseudopotentialstress}, while deriving $\sigma \mkern-2mu ^{E_{nl}}$---contribution to the stress tensor arising from the nonlocal pseudopotential energy---we have transferred the derivatives on the projectors (with respect to atomic position) to derivatives on the orbitals (with respect to space), as shown in Eqn.~\ref{Eqn:NonlocalReformulation}. This is because the orbitals are typically smoother than the projectors, and therefore the proposed strategy is expected to provide higher quality stresses, analogous to observations for the atomic forces \cite{hirose2005first,andrade2015real,ghosh2016sparc2}. To verify this, we consider an 8-atom unit cell of dc germanium and perform a $\Gamma$-point calculation with a mesh size of 0.44 Bohr. In Fig.~\ref{Fig:NLPcomparison}, we plot the convergence of the stress tensor with and without the reformulation of $\sigma \mkern-2mu ^{E_{nl}}$. It is clear from the results that the proposed formulation tremendously improves the accuracy of the stresses, and is therefore imperative to use in real-space DFT calculations. 

%%%%%%%%%%%%%%%%%%%%%%%%%%%%%%%%%%%%%%%%%%%%%%%%%%%%%%%%%%%%%%%%%%%%%%%%%%%%%%%%%%%%%%%%%%%%%%%%%%%%%%%%%%%%%

\section{On the stress tensor contribution $\sigma^{E_{self}}$ in real-space calculations}\label{subsection:stresswithoutselfenergycontribution}
\begin{figure}[h!]
\includegraphics[keepaspectratio=true,scale=0.5]{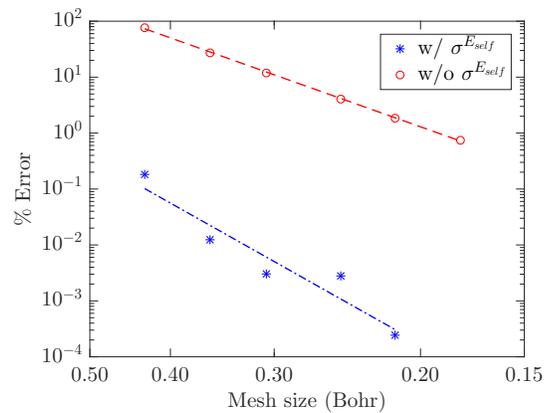}
\caption{\label{Fig:Selfcomparison} Convergence of the stress tensor with mesh size for the dc germanium system with and without the contribution arising from the self energy. The error is defined to be magnitude of the maximum difference in any component. The straight lines represent linear fits to the data.}
\end{figure}

As discussed in Section~\ref{Subsection:electrostaticstress}, it can be shown analytically that $\sigma^{E_{self}}=0$. However, due to the inexact nature of the chain rule within the finite-difference approximation, it is identically zero only in the limit of an infinitely fine mesh. In order to demonstrate the significant contribution of $\sigma^{E_{self}}$ in practical calculations, we consider an 8-atom unit cell of dc germanium and perform a $\Gamma$-point calculation with a mesh size of 0.44 Bohr. In Fig.~\ref{Fig:Selfcomparison}, we plot the convergence of the stress tensor with and without the contribution of $\sigma^{E_{self}}$. It is evident from the results that the proposed formulation of $\sigma^{E_{self}}$ tremendously improves the accuracy of the stresses, and is therefore imperative to use in real-space DFT calculations. 

%\bibliography{stress}  

\end{document}